\documentclass[12pt]{article}
\usepackage{amsmath,graphicx}
\usepackage{amssymb,hyperref}

\font\mybb=msbm10 at 12pt
\def\bb#1{\hbox{\mybb#1}}
\def\Z {\bb{Z}}
\def\R {\bb{R}}

\def\cF {{\cal F}}
\def\cO {{\cal O}}
\def\cN {{\cal N}}

\def\tr{\mathop{\rm tr}\nolimits}

\def\diag{\mathop{\rm diag}\nolimits}
\def\mod{\mathop{\rm mod}\nolimits}

\makeatletter
\@addtoreset{equation}{section}
\@addtoreset{footnote}{section}
\makeatother

\newcommand{\be}{\begin{equation}}
\newcommand{\ee}{\end{equation}}
\newcommand{\wt}{\widetilde}
\newcommand{\wh}{\widehat}
\newcommand{\ol}{\overline}

\newcommand{\ra}{\rightarrow}

\newcommand{\lra}{\leftrightarrow}
\newcommand{\nn}{\nonumber}
\newcommand{\half}{\frac{1}{2}}
\newcommand{\del}{\partial}

\def\half{\frac{1}{2}}

\newcommand{\f}{\frac}

\def \bea {\begin{eqnarray}}
\def \eea {\end{eqnarray}}
\def \beal#1 {\begin{align}#1\end{align}}
\def \nN {\notag\\}

\def\mat#1{\matt[#1]}
\def\matt[#1,#2,#3,#4]{\left(%
\begin{array}{cc} #1 & #2 \\ #3 & #4 \end{array} \right)}

\begin{document}

\begin{flushright}
\hfill{YITP-17-134}
\end{flushright}
\begin{center}
\vspace{2ex}
{\Large {\bf 
Deformation of ${\cal N}=4$ SYM with varying couplings
\\
\vspace{1ex}
via fluxes and intersecting branes
}}

\vspace*{5mm}
{\sc Jaewang Choi}$^{a}$\footnote{e-mail:
 {\tt jchoi@yukawa.kyoto-u.ac.jp}},
{\sc Jos\'e J. Fern\'andez-Melgarejo}$^{a,b}$\footnote{e-mail:
 {\tt josejuan@yukawa.kyoto-u.ac.jp}}
~and~
{\sc Shigeki Sugimoto}$^{a,c}$\footnote{e-mail:
 {\tt sugimoto@yukawa.kyoto-u.ac.jp}}

\vspace*{4mm} 

\hspace{-0.5cm}
{\it {$^{a}$
Center for Gravitational Physics, Yukawa Institute for Theoretical
 Physics,\\ Kyoto University, Kyoto 606-8502, Japan
}}

\vspace*{1mm}

{\it {$^{b}$
Departamento de F\'isica, Universidad de Murcia,\\
Campus de Espinardo,  30100 Murcia, Spain
}}

\vspace*{1mm}

{\it {$^{c}$
Kavli Institute for the Physics and Mathematics of the Universe (WPI),\\
 The University of Tokyo, Kashiwanoha, Kashiwa 277-8583, Japan
}}\\ 

\end{center}

\vspace*{.3cm}
\begin{center}
{\bf Abstract}
\end{center}

We study deformations of
${\cal N}=4$ supersymmetric Yang-Mills theory
with space-time dependent couplings by
embedding probe D3-branes in supergravity backgrounds
with non-trivial fluxes. The effective action on the world-volume
of the D3-branes is analyzed and a map between
the deformation parameters and the fluxes is obtained.
As an explicit example, we consider
D3-branes in a background corresponding to $(p,q)$ 5-branes
intersecting them and show that the effective theory
on the D3-branes precisely agrees with
the supersymmetric Janus configuration found by Gaiotto and Witten
in \cite{Gaiotto:2008sd}. D3-branes in an intersecting D3-brane
background is also analyzed and the D3-brane effective action
reproduces one of the supersymmetric configurations
with $ISO(1,1)\times SO(2)\times SO(4)$ symmetry
found in our previous paper \cite{Choi:2017kxf}.

\thispagestyle{empty}

\newpage

\tableofcontents

\section{Introduction}

As it is well-known, the effective theory on D3-branes
in flat space-time
becomes ${\cal N}=4$ supersymmetric Yang-Mills (SYM) theory
in the field theory limit ($\alpha'\ra 0$).
If the background has non-trivial fluxes, the effective theory
on the D3-branes will be deformed accordingly.
This is one of the useful ways to obtain 4 dimensional gauge theories
with less (or no) supersymmetry (SUSY).
In fact, various deformations realized in this way
have been investigated, for instance,
in \cite{Camara:2003ku,Grana:2003ek,Burgess:2006mn,McGuirk:2012sb}
in the context of flux compactifications.
In these works, because the main motivation
was to obtain a model beyond the Standard Model, the deformations
were assumed to preserve 4 dimensional Poincar\'e symmetry $ISO(1,3)$.
One of the main purposes of this paper is to generalize the deformations
to the cases where $ISO(1,3)$ is explicitly broken. In particular,
the couplings\footnote{In this paper, all the parameters, such as
gauge couplings, Yukawa couplings, theta parameter, masses, etc., in the
action are called ``couplings''.}
in the action may depend on the space-time coordinates.

In our recent paper \cite{Choi:2017kxf},
we wrote down the conditions to preserve part of the supersymmetry
in deformed ${\cal N}=4$ SYM with varying couplings
and found various non-trivial solutions.\footnote{
See,{\it e.g.}, \cite{Kim:2008dj,Gaiotto:2008sd,Kim:2009wv,Maxfield:2016lok}
for closely related works.
}
Though the motivation was in string theory, the analyses in \cite{Choi:2017kxf}
were purely field theoretical.
In this paper, we try to realize such systems in string theory
by putting probe D3-branes in supergravity backgrounds with fluxes and
find a map between the couplings in the action of the deformed
${\cal N}=4$ SYM and the fluxes in the background.

One way to obtain a theory with varying couplings is to consider
a background corresponding to D-branes (or other branes) intersecting
with the probe D3-branes. A typical example is a system with D3-branes
embedded in a background with $[p,q]$ 7-branes that appear as codimension
2 defects in the D3-brane world-volume \cite{Harvey:2007ab,Buchbinder:2007ar,
Harvey:2008zz}.\footnote{
See also \cite{Martucci:2014ema,Lawrie:2016axq,Couzens:2017way,Choi:2017kxf}
for recent related works.}
In this system, it is known that the complex coupling
(\ref{tau}) is not a constant but depends holomorphically on a complex
coordinate which is a complex combination of 2 spatial coordinates
transverse to the 7-branes.
Our results for the D3-brane effective action can be applied to this
system as well as various other intersecting brane systems. We demonstrate
it in two explicit examples;
intersecting D3-$(p,q)$5-brane and D3-D3 systems. In the former example,
we show that the effective action on the probe D3-branes
precisely reproduces the action for the supersymmetric Janus configuration
found in \cite{Gaiotto:2008sd}.
The latter example reproduces one of the supersymmetric solutions
with $ISO(1,1)\times SO(2)\times SO(4)$
symmetry obtained in \cite{Choi:2017kxf}.

The contents of the paper is as follows.
In Section \ref{sec:SYM}, we review the deformations of ${\cal N}=4$ SYM
with varying couplings that were studied in \cite{Choi:2017kxf}.
In particular, we summarize the SUSY conditions and a few explicit solutions
that will be used in Section \ref{sec:examples}.
In Section \ref{sec:D3-curved}, we study the effective action of a stack of
D3-branes in curved backgrounds with fluxes.
Upon the leading order expansion with respect to $\alpha'$,
we establish a map between the background fields and the deformation
parameters of the theory in Section \ref{sec:SYM}.
In Section \ref{sec:examples}, we apply the
results of the previous section to study two cases: backgrounds with
$(p,q)$ 5-branes and those with D3-branes. These two examples
correspond to the realization of the deformed ${\cal N}=4$ SYM
studied in \cite{Gaiotto:2008sd} and \cite{Choi:2017kxf} in string theory.
In Section \ref{sec:conclusions}, we conclude the paper with various discussions
on the present work and further applications. In addition, two appendices
are included. In Appendix \ref{app:SUGRA}, we summarize our
conventions used for the supergravity fields.
In Appendix \ref{app:D3action}, we show the explicit calculations to obtain
the effective action of D3-branes given in Section \ref{sec:D3-curved}.

\section{Deformations of ${\cal N}=4$ SYM with varying couplings}
\label{sec:SYM}

In this section, we review some of the results obtained
in \cite{Choi:2017kxf}. Following \cite{Choi:2017kxf},
we use a 10-dimensional
notation, in which ${\cal N}=4$ SYM
is regarded as a dimensional reduction of 10-dimensional ${\cal N}=1$ SYM.
The 10-dimensional gauge field $A_I$ ($I=0,\dots,9$) is reduced to
a 4-dimensional gauge field $A_\mu$ ($\mu=0,\dots,3$) and 6 scalar
fields $A_A$ ($A=4,\dots,9$), and the 10-dimensional Majorana-Weyl
spinor field $\Psi$ describes 4 Weyl fermion fields in 4-dimensions.
$\Psi$ is a 32-component Majorana spinor satisfying the Weyl condition
\begin{align}
\Gamma^{(10)}\Psi = +\Psi
\ ,
\label{chiralcond}
\end{align}
where $\Gamma^{(10)}\equiv \Gamma^{\hat0} \dots \Gamma^{\hat9}$ is the
10-dimensional chirality operator.\footnote{
In \cite{Choi:2017kxf}, the fermions are chosen to have negative chirality
(minus sign in the right hand side of (\ref{chiralcond})). Here,
we choose the chirality to be positive, in order to match the convention
used in \cite{Martucci:2005rb}. One way to relate our convention here and
that in \cite{Choi:2017kxf} is to use a transformation $x^9\ra-x^9$,
which induces $\Psi^{\rm here}=\Gamma^{\hat 9}\Psi^{\rm there}$,
$A_9^{\rm here}=-A_9^{\rm there}$,
$A_{I'}^{\rm here}=A_{I'}^{\rm there}$ for $I'\ne 9$, and
similar sign changes for the parameters $d^{IJA}$, $m^{AB}$
and $m_{IJK}$.
} The gamma matrices $\Gamma^{\hat I}$ ($\hat I=0,\dots,9$)
are 10-dimensional gamma matrices which are realized as $32\times32$
real matrices satisfying $\{\Gamma^{\hat I}, \Gamma^{\hat J}\}
= 2\eta^{\hat I\hat J}$, where $\eta^{\hat I \hat J} =\diag (-1,+1,\dots,+1)$
is the 10-dimensional Minkowski metric.

Let us consider the following deformation of
the ${\cal N}=4$ $SU(N)$ supersymmetric Yang-Mills theory:
\begin{multline}
 S=\int d^4 x \sqrt{-g}\,a\tr
\bigg\{
-\frac{1}{2}\,g^{II'}g^{JJ'}F_{IJ}F_{I'J'}+
i\,\ol\Psi\Gamma^I D_I\Psi
+\frac{c}{4}\,\epsilon^{\mu\nu\rho\sigma}F_{\mu\nu}F_{\rho\sigma}
\\
-d^{IJA}F_{IJ}A_A-\frac{m^{AB}}{2}A_AA_B
-i\ol\Psi M \Psi
\bigg\}
\ ,
\label{action1}
\end{multline}
where $I,J=0,\dots,9$; $\mu,\nu=0,\dots,3$; $A,B=4,\dots,9$.
$F_{IJ}$ is defined as
\begin{align}
&F_{\mu\nu}
\equiv
\del_\mu A_\nu-\del_\nu A_\mu + i[A_\mu,A_\nu]
\ ,
\\
&F_{\mu A}=-F_{A\mu} 
\equiv
\partial_{\mu} A_{A} + i[A_\mu,A_A]
\equiv 
D_\mu A_A
\ ,
\\
&F_{AB}
\equiv
i[A_A,A_B]
\ .
\end{align}
The covariant derivatives on the fermion field $\Psi$ are defined as
\begin{align}
D_\mu \Psi
\equiv\
\partial_\mu\Psi +i[A_\mu,\Psi]
+\frac14 \omega_{\mu}^{~\hat \nu\hat\rho} \Gamma_{\hat \nu\hat\rho} \Psi
\ ,
~~~~
D_A\Psi \equiv\ i[A_A,\Psi]
\ ,
\label{delPsi}
\end{align}
where the indices  $\hat\mu,\hat \nu=0,\ldots,3$ are flat indices,
$\Gamma_{\hat\nu\hat\rho}\equiv \half(\Gamma_{\hat\nu}\Gamma_{\hat\rho}
-\Gamma_{\hat\rho}\Gamma_{\hat\nu})$
and
$\omega_{\mu}^{~\hat\nu\hat\rho}$ is the spin connection.
We assume that the metric  $g_{IJ}$ has the form
\begin{align}
ds^2
=
g_{IJ} dx^I dx^J
=
g_{\mu\nu}(x^\rho)
dx^\mu dx^\nu
+
\delta_{AB} dx^A dx^B
\, ,
\label{metricSYM}
\end{align}
and $g^{IJ}$ denotes its inverse. The 4-dimensional 
Levi-Civita symbol $\epsilon^{\mu\nu\rho\sigma}$ is defined such that
$\epsilon^{0123}=1/\sqrt{-g}$, where
$\sqrt{-g}\equiv \sqrt{-\det(g_{\mu\nu})}$.
We also introduce a vielbein $e_I^{\hat I}$
satisfying $e_I^{\hat I}e_J^{\hat J}\eta_{\hat I\hat J}=g_{IJ}$
and its inverse $e^I_{\hat I}$. The gamma matrices with the curved
indices are defined by $\Gamma^I=e^I_{\hat I}\Gamma^{\hat I}$.

The quantities $a$, $c$, $d^{IJA}$, $m^{AB}$ are real
parameters and $M$ is a $32\times 32$ real anti-symmetric
matrix. All of them may depend on the space-time coordinates $x^\mu$.
In this paper, we call these parameters as ``couplings'',
though $m^{AB}$ and $M$ are related to masses.
The couplings $a$ and $c$ are related to the gauge coupling $g_{\text{YM}}$
and the theta parameter $\theta$ as follows:
\begin{align}
a=\frac{1}{g_{\text{YM}}^2}
\ ,
\qquad
c=\frac{g_{\text{YM}}^2 \theta }{8\pi^2}
\ .
\end{align}
It is useful to define the complex coupling $\tau$
in terms of these quantities:
\begin{align}
\tau
\equiv
\frac{\theta}{2\pi}+i \frac{4\pi}{g_{\text{YM}}^2}
=4\pi a(c+i)
\ .
\label{tau}
\end{align}

The parameters $d^{IJA}$ and $m^{AB}$ exhibit the following symmetries
\begin{eqnarray}
 d^{IJ A}= -d^{JI A}\ ,
 \qquad
 d^{\mu A B}= -d^{\mu BA}\ ,
 \qquad
 d^{A B C}= d^{[ABC]}\ ,
 \qquad
 m^{AB}=m^{BA}\ ,
\label{d-and-m}
\end{eqnarray}
whereas the most general form for $M$ is given by
\begin{align}
M= m_{IJK} \Gamma^{IJK}
\ ,
\end{align}
where $m_{IJK}$ is a real rank-3 anti-symmetric tensor
and
\begin{align}
\Gamma^{IJK}\equiv&\ 
\Gamma^{[I}\Gamma^J\Gamma^{K]}
\nn\\
\equiv&\
\frac{1}{3!}(
\Gamma^I\Gamma^J\Gamma^K+\Gamma^J\Gamma^K\Gamma^I
+\Gamma^K\Gamma^I\Gamma^J
-\Gamma^J\Gamma^I\Gamma^K-\Gamma^I\Gamma^K\Gamma^J-\Gamma^K\Gamma^J\Gamma^I)\ .
\end{align}

The ansatz for the SUSY transformation is
\begin{eqnarray}
\delta_\epsilon A_I=i\ol\epsilon\,\Gamma_I\Psi\ ,
~~~
\delta_\epsilon \ol\Psi=
\frac{1}{2}\ol\epsilon\,(-F_{IJ}\Gamma^{IJ}+A_A\ol B^A)\ ,
\end{eqnarray}
where $\epsilon$ is the SUSY parameter represented
as a 10-dimensional Majorana-Weyl spinor and
 $\ol B^A$ is a $32\times 32$ real matrix.
Both $\epsilon$ and $\ol B^A$ may depend on space-time.

Then, the invariance of the action \eqref{action1} under this
SUSY transformation implies the following equations:
\begin{align}
0
=&\
\ol\epsilon\, e^{I'J'K'}
\Gamma_{K'}
\left(
\frac{1}{72}\Gamma_{I'J'}\Gamma^{IJ}
+\frac{1}{4}\Gamma_{[I'}\Gamma^{[I}\delta_{J']}^{J]}
-\delta_{I'}^I\delta_{J'}^J
\right)\ ,
\label{SUSYcond1}
\\
0 
=&\
\ol\epsilon\,\bigg(\frac{1}{72}e^{IJK}\Gamma_{IJK}
-
\frac{1}{2}\Gamma^{\mu}\del_\mu\log a
-\left(
\frac{1}{16}e_{\mu JK}-3m_{\mu JK}\right)\Gamma^{\mu JK}
-M
\bigg)\ ,
\\
\ol\epsilon\,
\ol B^{A}
=&\ 
\ol\epsilon\left(F\Gamma^A+
\left(
-\frac{1}{4}e^{AJK}+12 m^{AJK}
\right)\Gamma_{JK}
\right)
\ ,
\\
\del_\mu\ol\epsilon
=&\
-\frac{1}{4}\ol\epsilon\left(
F\Gamma_\mu+\left(
-\frac{1}{4}e_{\mu\hat J\hat K}
+12m_{\mu\hat J\hat K}-\omega_{\mu\hat J\hat K}
\right)\Gamma^{\hat J\hat K}
\right)
\ ,
\\
D_\mu (\ol\epsilon\,\ol{B}^{A})\Gamma^\mu
=&\
\ol\epsilon
\left(-2a^{-1}\Gamma_ID_\mu (a\,d^{I\mu A})-m^{AB}\Gamma_B 
-\ol{B}^A\left(M+\half\Gamma^\mu\del_\mu\log a\right)
\right)
\ ,
\label{SUSYcond5}
\end{align}
where $F$ is a real $32\times 32$ matrix acting on the spinor
indices and
\begin{eqnarray}
e^{IJK}\equiv a^{-1}\del_\nu(ac)\,\epsilon^{\nu IJK}
+3\,d^{[IJK]}+24\,m^{IJK}\ .
\end{eqnarray}
The condition (\ref{SUSYcond1}) has a trivial solution,  $e^{IJK}=0$,
which is equivalent to
\begin{align}
0=&\ a^{-1}\del_\mu(ac)\,\epsilon^{\mu\nu\rho\sigma}
+24\,m^{\nu\rho\sigma}\ ,
\label{delac}
\\
0=&\ d^{[IJA]}+8\,m^{IJA}
\ .
\end{align}
Using the symmetries of the deformation parameters (\ref{d-and-m}), the
latter is written as
\begin{eqnarray}
d^{\mu\nu A}= -24\,m^{\mu\nu A} 
\ ,
\qquad
d^{\mu AB}=-12\,m^{\mu AB} 
\ ,
\qquad
 d^{ABC}=-8\,m^{ABC}\ .
\label{md}
\end{eqnarray}

Further discussions on the nature of these equations and their
solutions, we refer to \cite{Choi:2017kxf}. Let us summarize
a few explicit solutions that are relevant for our discussion.

\begin{enumerate}

\item $ISO(1,2)\times SO(3)\times SO(3)$

The case with $ISO(1,2)\times SO(3)\times SO(3)$ symmetry is analyzed
by Gaiotto and Witten in \cite{Gaiotto:2008sd}. (See also
section 3.4 in \cite{Choi:2017kxf}) It is a solution of
the SUSY conditions (\ref{SUSYcond1})--(\ref{SUSYcond5}) with
the parameters depending only on $x^3$.
$ISO(1,2)$ is the Poincar\'e group acting on $x^{0,1,2}$ and
$SO(3)\times SO(3)$ acts on $x^{4,5,6}$ and $x^{7,8,9}$.
The metric is assumed to be flat and the non-trivial components of
the couplings in the action are given as follows:\footnote{
Our convention is slightly different from that in \cite{Gaiotto:2008sd}.
The solution shown here is taken from section 3.4
in \cite{Choi:2017kxf} with $b_0=2$ and $l(z)=-iz/\sqrt{2}$.
We also made a transformation $x^9\ra -x^9$. (See the footnote in p.3.)
}
\begin{align}
&\tau= 4\pi a(c+i)=\tau_0+4\pi D\,e^{i2\psi}\ ,
\label{tauGW}
\\
&m^{ab}=2\left(\psi'^2-(\psi'\cot\psi)'\right)\delta^{ab}
\ ,~~~(a,b=4,5,6)\nn\\
&
m^{pq}=2\left(\psi'^2+(\psi'\tan\psi)'\right)\delta^{pq}
\ ,~~~(p,q=7,8,9)
\label{mGW}
\\
&d^{456}=\frac{2}{3}\frac{\psi'}{\sin\psi}\ ,~~~
d^{789}=\frac{2}{3}\frac{\psi'}{\cos\psi}\ ,
\label{dGW}
\\
&M=\frac{\psi'}{2}\Gamma^{012}-\frac{\psi'}{2\sin\psi}\Gamma^{456}
-\frac{\psi'}{2\cos\psi}\Gamma^{789}\ ,
\label{MfermiGW}
\end{align}
where
 $\tau_0$ and $D$ are real constants and $\psi$ is an arbitrary
real function of $x^3$ with $0<\psi<\pi/2$ assuming $D>0$.

\item $ISO(1,1)\times SO(2)\times SO(4)$

The case with $ISO(1,1)\times SO(2)\times SO(4)$ symmetry
is given in section 4.1 in \cite{Choi:2017kxf}. Here, $ISO(1,1)$,
$SO(2)$ and $SO(4)$ act on $x^{0,1}$, $x^{4,5}$ and $x^{6,7,8,9}$,
respectively, and the couplings in the action may depend on $x^{2,3}$.
The metric (\ref{metricSYM}) is assumed to be
\begin{eqnarray}
 ds^2=\eta_{\alpha\beta}dx^\alpha dx^\beta+e^{\varphi}\delta_{mn}dx^mdx^n
+\delta_{ab}dx^adx^b+\delta_{pq}dx^pdx^q\ ,
\label{2-4metric}
\end{eqnarray}
where the indices are $\alpha,\beta=0,1$; $m,n=2,3$;
$a,b=4,5$ and $p,q=6,7,8,9$.
In this case, the complex coupling (\ref{tau}) turns out to be
an arbitrary holomorphic (or anti-holomorphic) function of
a complex coordinate $z\equiv\frac{1}{\sqrt{2}}(x^2+ix^3)$ with ${\rm Im}\,\tau>0$
and $\varphi$ in the metric (\ref{2-4metric}) is an arbitrary
real function of $x^{2,3}$.
$M$ is of the form:
\begin{eqnarray}
 M=\alpha_m\Gamma^{01m}+\beta_m\Gamma^{m45}\ ,
\end{eqnarray}
and $\alpha_m$ and $\beta_m$ are determined by $\tau$ and $\varphi$ as
\begin{align}
\alpha_m=&\ \frac{1}{4}\del_m\log{\rm Im}\,\tau
=\frac{1}{4}({\rm Im}\,\tau)^{-1}\del_n({\rm Re}\,\tau)
\,\epsilon^{n}_{~m}\ ,
\\
\beta_m=&\ \frac{s}{4}\epsilon^n_{~m}\del_n\left(
\varphi-\log{\rm Im}\,\tau
\right)+\del_m\Lambda\ ,
\label{beta2-4}
\end{align}
where $s=\pm$,
 $\epsilon^{nm}=\epsilon^n_{~m'}g^{m'm}$ is the
Levi-Civita symbol for the $x^{2,3}$-plane and $\Lambda$
is an arbitrary real function.\footnote{$\Lambda$ can be absorbed
by a local $SO(2)$ rotation of the $x^{4,5}$-plane.
See Appendix C.2 in \cite{Choi:2017kxf}.}
The non-trivial components of $d^{IJA}$ and $m^{AB}$ are
\begin{eqnarray}
&& d^{nab}=- d^{anb}=-2\beta^n\epsilon^{ab}\ ,
\label{dnab2-4}
\\
&&m^{ab}=\left(-\half g^{mn}q_mq_n-g^{mn}\del_mq_n+8g^{mn}\beta_m\beta_n
-4s\del_m\beta_n\epsilon^{mn}\right)\delta^{ab}\ ,
\label{mab2-4}
\\
&&m^{pq}=\left(-\half g^{mn}q_mq_n-g^{mn}\del_mq_n\right)\delta^{pq}\ ,
\label{mpq2-4}
\end{eqnarray}
where
\begin{eqnarray}
q_m\equiv \del_m\log{\rm Im}\,\tau= \del_m\log a\ .
\end{eqnarray}

\item $ISO(1,1)\times SO(6)$

When $\beta_m=0$ in the previous example, the symmetry
is enhanced to $ISO(1,1)\times SO(6)$. In this case,
$(\varphi-\log{\rm Im}\,\tau)$ is a harmonic function
on $x^{2,3}$-plane satisfying
\begin{eqnarray}
g^{mn}\del_m\del_n\left(\varphi-\log{\rm Im}\,\tau\right) =0\ .
\end{eqnarray}
This is the case studied in
\cite{Harvey:2007ab,Buchbinder:2007ar,Harvey:2008zz}.
(See also section 3.3 in \cite{Choi:2017kxf})
It is related to the effective theory on the D3-branes
embedded in a 7-brane background as mentioned in the introduction.

\end{enumerate}

\section{D3-branes in curved backgrounds with fluxes}
\label{sec:D3-curved}

In this section, we study the effective action of
D3-branes in curved backgrounds with fluxes
and try to relate the couplings
in the action (\ref{action1}) with the supergravity fields.
As reviewed in Appendix \ref{Dpreview},
the effective action of D$p$-branes in general backgrounds
is known, at least, to the extent needed for our purpose.
(See (\ref{bosonic-initial}) and (\ref{Sfermi}) for the bosonic
and fermionic parts, respectively.)
However, the expression of the effective action reviewed in Appendix
\ref{Dpreview} is not convenient for a direct comparison with the action
(\ref{action1}) used in the field theoretical analysis.

To find a relations between couplings in (\ref{action1}) and the
fluxes in the supergravity background,
we expand the D3-brane effective action with respect
to $\alpha'=l_s^2$ and keep only the terms that survive in the
$\alpha'\ra 0$ limit, assuming that the background fields are
of $\cO(\alpha'^0)$.

We consider $N$ D3-branes embedded in a 10 dimensional space-time
parametrized by $(x^\mu, x^i)$ with $\mu=0,1,2,3$ and $i=4,\dots,9$.
We use the static gauge, in which the world-volume of the D3-branes
is parametrized by $x^\mu$ ($\mu=0,1,2,3$). The scalar fields, which are
related to $A_A$ ($A=4,\dots,9$) in the previous section,
are denoted here as $\Phi^i$ ($i=4,\dots,9$).
The scalar field $\Phi^i$ describes the position of the D3-branes
in the $x^i$ direction. Assuming that the D3-branes are placed at $x^i=0$
when $\Phi^i=0$, the relation between the position of D3-branes
and the value of scalar fields is given by
$\Phi^i=\lambda x^i$ with
$\lambda\equiv 2\pi\alpha'=2\pi l_s^2$. (See (\ref{expand}) for
the precise meaning of this identification for $N>1$.)

To simplify the analysis, we assume that $(\mu,i)$ components
of the metric $g_{\mu i}$ vanish everywhere, and all
the components of the Kalb-Ramond 2-form fields and all the R-R fields,
except the R-R 0-form $C_0$, vanish at $x^i=0$
($i=4,\dots,9$)\footnote{
It is generically possible to choose a gauge such that
$B_2|_{x^i=0}= 0$ and $C_{n}|_{x^i=0}=0$ $(n\ne 0)$ (at least locally)
provided the components $H_{\mu\nu\rho}$ and $F_{\mu\nu\rho}$
vanish at $x^i=0$. Obviously, the reason for
considering a non-vanishing $C_0$ is that we want to capture the
theta parameter $\theta$ in the SYM action \eqref{action1}.}:
\begin{align}
g_{\mu i}= 0\ ,~~~
B_2\big|_{x^i=0} = 0\ ,~~~
C_{n}\big|_{x^i=0}= 0
\ ,
~~(n\neq 0)
\ .
\label{assumptions}
\end{align}
Note that unlike in (\ref{metricSYM}),
the $(i,j)$ component of the metric $g_{ij}$ may have non-trivial
$x^\mu$ dependence.

Here, we simply state our results on the D3-brane effective action
and leave the details to Appendix \ref{app:D3action}.
Neglecting the $\cO(\alpha')$ terms in the action
(\ref{bosonic-initial}) and (\ref{Sfermi}),
we obtain:
\begin{align}
S_{\rm D3}^{\rm boson}
=&\
\frac{T_3\lambda^2}{2}\int d^4 x
\sqrt{-g}
\tr\bigg\{
-\frac{e^{-\phi}}{2}
 g^{\mu\nu}g^{\rho\sigma}F_{\mu\rho}F_{\nu\sigma}
\pm\frac{C_0}{4}
\epsilon^{\mu\nu\rho\sigma}F_{\mu\nu}F_{\rho\sigma}
\nn\\
&
\qquad\qquad
-e^{-\phi}
 g^{\mu\nu}g_{ij}D_\mu\Phi^i D_\nu\Phi^j
+\frac{e^{-\phi}}{2} g_{ii'}g_{jj'}[\Phi^i,\Phi^j][\Phi^{i'},\Phi^{j'}]
-2V(\Phi^i)
\nn\\
&
\qquad\qquad
\pm (G^R_\pm)_i^{~\mu\nu}
\Phi^iF_{\mu\nu}
\mp\frac{i}{3}(G^R_\pm)_{ijk}\Phi^i[\Phi^j,\Phi^k]
\mp (*_4\wt F_5)^\mu_{~ij}\Phi^iD_\mu\Phi^j
\bigg\}\ ,
\label{SboseD3}
\\
 S_{\rm D3}^{\rm fermi}
=&\
\frac{T_3\lambda^2}{2}\int d^4 x\, \sqrt{-g}\,e^{-\phi}\tr \left\{
i\left(\ol\Psi\Gamma^\mu D_\mu\Psi
+\ol\Psi\Gamma_k i[\Phi^k,\Psi]\right)
-i\ol\Psi\left( M_\pm
-\frac{1}{4}\omega_{\mu\hat i\hat j}\Gamma^{\mu\hat i\hat j}
\right) \Psi \right\}\ ,
\label{SfermiD3}
\end{align}
where the upper (lower) signs correspond to the case of 
D3- ($\overline{\text{D}3}$-) branes,
 $D_\mu$ denotes
the 4 dimensional covariant derivative defined in (\ref{delPsi})
and (\ref{delPhi}), and $\omega_{\mu\hat i\hat j}$ is the $(\mu,\hat i,\hat j)$
component of the spin connection\footnote{
The hatted indices are the flat indices as in the previous section.
We assume $e^i_{\hat \mu}=0$ and $e^\mu_{\hat i}=0$ without loss of
generality under the assumption (\ref{assumptions}).}
related to the vielbein $e^i_{\hat j}$ as
\begin{eqnarray}
\omega_{\mu \hat i\hat j}
=\frac{1}{2}
(e^k_{\hat i}\del_\mu e_{k\hat j}-e^k_{\hat j}\del_\mu e_{k\hat i})\ .
\label{omega-muij}
\end{eqnarray}

The fluxes $(G_\pm^R)_{i}^{~\mu\nu}$, $(G_\pm^R)_{ijk}$
and $(*_4\wt F_5)^\mu_{~ij}$ in (\ref{SboseD3})
are defined in (\ref{GR1}), (\ref{GR2}) and (\ref{wtF}).
See also Appendix \ref{app:SUGRA} for our conventions
for the supergravity fields.

The quantity $M_\pm$ in \eqref{SfermiD3} is given by 
\begin{eqnarray}
M_\pm \equiv
\mp \frac{e^{\phi}}{8}\left(
\frac{1}{3}(*_4 F_1)_{\mu\nu\rho}\Gamma^{\mu\nu\rho}
- (G^R_\pm)_{i\mu\nu}\Gamma^{i\mu\nu}
+\frac{1}{3}\,(G_\pm^R)_{ijk}\Gamma^{ijk}
-(*_4\wt F_5)_{\mu ij}\Gamma^{\mu ij}
\right)
\ ,
\end{eqnarray}
where $(*_4F_1)_{\nu\rho\sigma}$ is defined in (\ref{F1star}).

The potential $V(\Phi^i)$ in \eqref{SboseD3} has two contributions:
\begin{eqnarray}
 V(\Phi^i)\equiv\lambda^{-2}(V_{\rm DBI}(\Phi^i)\pm V_{\rm CS}(\Phi^i))\ ,
\label{VPhi}
\end{eqnarray}
where
\begin{eqnarray}
V_{\rm DBI}(\Phi^i)&\equiv&
e^{-\phi}\left(
1+\lambda
v^{\rm DBI}_i\Phi^i
+\frac{\lambda^2}{2}m^{\rm DBI}_{ij}
\Phi^i\Phi^j
\right)
\ ,
\label{VDBI}
\\
V_{\rm CS}(\Phi^i)&\equiv&
\lambda
v^{\rm CS}_i\Phi^i
+\frac{\lambda^2}{2}m^{\rm CS}_{ij}
\Phi^i\Phi^j
\ ,
\end{eqnarray}
with
\begin{align}
v^{\rm DBI}_i
\equiv&\
-\del_i\phi+\frac{1}{2}g^{\mu\nu}\del_ig_{\mu\nu}
=\del_i\log\left(\sqrt{-g}\,e^{-\phi}\right)
\ ,
\\[4pt]
m^{\rm DBI}_{ij}
\equiv&\
v_i^{\rm DBI}v_j^{\rm DBI}
-\del_i\del_j\phi
+\half\left(
g^{\mu\nu}\del_i\del_j g_{\nu\mu}
-g^{\mu\mu'}\del_i g_{\mu'\nu'} g^{\nu'\nu}\del_j g_{\nu\mu}
-g^{\mu\mu'}g^{\nu\nu'}H_{i\mu'\nu'} H_{j\nu\mu}
\right)
\nn\\
=&\ \frac{1}{\sqrt{-g}\,e^{-\phi}}\del_i\del_j
\left(\sqrt{-g}\,e^{-\phi}\right)
+\half
H_{i}^{~\mu\nu} H_{j\mu\nu}\ ,
\\
v_i^{\rm CS}
 \equiv&\
- (*_4\wt F_5)_i\ ,
\\[4pt]
 m_{ij}^{\rm CS}
\equiv&\
-\frac{1}{2}
\left(\del_i(*_4\wt F_5)_{j}+\half(*_4\wt F_3)^{~\mu\nu}_{j}H_{\mu\nu i}
+(i\lra j)
\right)
\ ,
\end{align}
and $(*_4\wt F_5)_j$ and $(*_4\wt F_3)_j^{~\mu\nu}$ are defined in (\ref{wtF}).

All the supergravity fields and their derivatives
in the action (\ref{SboseD3}) and (\ref{SfermiD3})
are evaluated at $x^i=0$. The first term in $V_{\rm DBI}$
(\ref{VDBI})
can be discarded in the comparison with the field theory results,
because it doesn't depend on $\Phi^i$.
If we require $\Phi^i=0$ and $A_\mu=0$ to be a solution
of the equations of motion, the linear term in (\ref{VPhi})
has to vanish:
 \begin{eqnarray}
 0=e^{-\phi}v_i^{\rm DBI}\pm v_i^{\rm CS}\ ,
\label{linear}
\end{eqnarray}
which is the condition that the force due to NS-NS and R-R fields
cancel each other.
In this case, we can safely take the $l_s\ra 0$ limit.

Since the metric used in the action (\ref{action1}) is assumed to be
of the form (\ref{metricSYM}), we introduce a new metric
\begin{eqnarray}
\bar g_{\mu\nu}\equiv e^{2\omega} g_{\mu\nu}\ ,~~~
\bar g_{AB}\equiv \delta_{AB}\ ,~~~
\bar g_{\mu A}\equiv 0\ ,
\label{barmetric}
\end{eqnarray}
where $\mu,\nu=0,\dots,3$; $A,B=4,\dots,9$ and
$\omega$ is a real function. Here, we put a factor $e^{2\omega}$
in the 4-dimensional metric, because it is often convenient to make a
Weyl transformation to get a metric $\bar g_{IJ}$ that can be identified
with $g_{IJ}$ used in the previous section.\footnote{See Appendix C.1 in
\cite{Choi:2017kxf} for useful formulas for the Weyl transformation.}
In addition, we redefine the scalar fields as
\begin{eqnarray}
A^{A}\equiv e^{-\omega} e^{A}_{i} \Phi^i\ ,
\end{eqnarray}
where $e^A_i$ is the vielbein for the transverse space $e^{\hat i}_i$
with the identification $A=\hat i =4,\dots,9$.
so that the kinetic term can be written as in (\ref{action1}) with the metric
$\bar g_{IJ}$ defined in (\ref{barmetric}).

Then, discarding the total derivative terms,
the bosonic part of the D3 brane action $\eqref{SboseD3}$ becomes
\beal{
S_{\rm D3}^{\rm boson}
=&\
\frac{T_3\lambda^2}{2}\int d^4 x
\sqrt{-\bar{g} }
\tr\bigg\{
-\frac{e^{-\phi}}{2}\bar{g}^{IJ} \bar{g}^{KL}F_{IK}F_{JL}
\pm\frac{C_0}{4}
\bar\epsilon^{\mu\nu\rho\sigma}F_{\mu\nu}F_{\rho\sigma}
\pm e^{-3\omega} (G^R_\pm)_A^{~\mu\nu} A^A F_{\mu\nu}
\nn \\ 
&\ 
\quad\mp\frac{i}{3}e^{-\omega}(G^R_\pm)_{ABC} A^A [A^B,A^C]
+\left( \mp (*_4\wt F_5)_{\mu AB }
+2e^{-\phi}\omega_{\mu AB}\right)\bar g^{\mu\nu}
A^A D_\nu A^B
\nn\\
&\
\quad-\wh m_{AB}A^AA^B-2e^{-4\omega} V(\Phi^i)
\bigg\}\ ,
\label{Sboson-2}
}
where $\ol\epsilon^{\mu\nu\rho\sigma}$ is the Levi-Civita symbol
with $\ol\epsilon^{0123}=1/\sqrt{-\bar g}$,
supergravity fields with indices $A,B,C$ are defined as
$(G_\pm^R)_{ABC}\equiv (G_\pm^R)_{ijk}e^i_Ae^j_Be^k_C$,
etc., and $\wh m_{AB}$ is defined as
\begin{align}
\wh m_{AB}
\equiv
\half\bigg(
e^{-\phi}\bar g^{\mu\nu}E_{\mu A'A}E^{~A'}_{\nu~B}
-
\frac{1}{\sqrt{-\bar g}}\del_\nu\left(\sqrt{-\bar g}\,e^{-\phi}\bar g^{\mu\nu}
E_{\mu AB} \right)
\pm e^{-2\omega}(*_4\wt F_5)^\mu_{~AA'}E^{~A'}_{\mu~B}
\bigg)+(A\lra B)
\end{align}
with
\begin{eqnarray}
 E_{\mu AB}\equiv e^{-\omega}e_{iA}\del_\mu(e^\omega e^i_B)
=
\half \del_\mu g^{ij}e_{iA}e_{jB}+\del_\mu\omega\delta_{AB}+\omega_{\mu AB}
\ .
\end{eqnarray}

The fermionic part (\ref{SfermiD3}) is rewritten as
\begin{align}
S_{\rm D3}^{\rm fermi}
=&\
\frac{T_3\lambda^2}{2}\int d^4 x\, \sqrt{-\bar g}\,e^{-\phi}\tr \left\{
i(\ol{\wh\Psi}\wh\Gamma^\mu D_\mu{\wh\Psi}
+\ol{\wh\Psi}\wh\Gamma_A i[A^A,\wh\Psi])
-i\ol{\wh\Psi}\left( \wh M_\pm
-\frac{1}{4}\omega_{\mu AB}\wh\Gamma^{\mu AB}
\right)\wh\Psi \right\}\ ,
\nn\\
\label{Sfermi-2}
\end{align}
where we have defined
\begin{align}
\wh M_\pm \equiv&\
e^{-\omega}M_\pm
\nn\\
=&\
\mp \frac{e^{\phi}}{8}\left(
\frac{e^{2\omega}}{3}(*_4 F_1)_{\mu\nu\rho}\wh\Gamma^{\mu\nu\rho}
-e^{\omega} (G^R_\pm)_{A\mu\nu}\wh\Gamma^{A\mu\nu}
+\frac{e^{-\omega}}{3}\,(G_\pm^R)_{ABC}\wh\Gamma^{ABC}
-(*_4\wt F_5)_{\mu AB}\wh\Gamma^{\mu AB}
\right)
\ ,
\end{align}
and
\begin{align}
\wh\Gamma^\mu\equiv e^{-\omega}\Gamma^\mu\ ,~~~
\wh\Gamma^{A}\equiv e^{A}_i\Gamma^{i}\ ,~~~
\wh\Psi\equiv e^{-\frac32\omega}\Psi\ .
\end{align}
Here, $\wh\Gamma^I$ ($I=0,1,\dots,9$)
are the gamma matrices satisfying
\begin{eqnarray}
 \{\wh\Gamma^I,\wh\Gamma^J\}=2\bar g^{IJ}\ .
\end{eqnarray}

Now, we can readily find the correspondence between the couplings
and the supergravity fields.
By comparing the action (\ref{action1})
with (\ref{Sboson-2}) and (\ref{Sfermi-2}), assuming (\ref{linear}),
we obtain
\begin{align}
&
a = \frac{T_3\lambda^2}{2}e^{-\phi}\ , \quad
c= \pm e^\phi C_0\ ,\quad
\label{map_first}
d^{\mu\nu A}
=\mp e^{-3\omega+\phi} (G_{\pm}^R)^{A\mu\nu}\  ,
\\
&
d_{\mu AB}
=  \mp \f{e^{\phi}}{2}  (*_4\wt F_5)_{\mu AB}
+\omega_{\mu AB} \ ,\quad
d^{ABC}
=\pm\frac{e^{-\omega+\phi}}{3}(G_{\pm}^R)_{ABC}
\ , \label{map_second}  \\
&
m_{\mu\nu\rho}=\mp\frac{e^{2\omega+\phi}}{24}(*_4 F_1)_{\mu\nu\rho}
\ ,\quad
m_{\mu\nu A}
=\pm\frac{e^{\omega+\phi}}{24}(G_\pm^R)_{A\mu\nu}\ ,
\label{map_third}
\\
&
m_{\mu AB}
=\pm\frac{e^{\phi}}{24}(*_4\wt F_5)_{\mu AB} -\f{1}{12}
\omega_{\mu AB}\ ,\quad
m_{ABC}
=\mp\frac{e^{-\omega+\phi}}{24}(G_\pm^R)_{ABC}\ ,
\label{map_fourth}
\end{align}
and
\begin{eqnarray}
m_{AB}
=
2(\wh m_{AB}+e^{-2\omega}m_{AB}^{\rm DBI}
\pm e^{-2\omega+\phi} m_{AB}^{\rm CS})
\ .
\label{map_last}
\end{eqnarray}

Note that the first equation of (\ref{map_third}) can be written as
\begin{eqnarray}
m_{\mu\nu\rho}
=
\mp\frac{e^{2\omega+\phi}}{24}
 \epsilon^{\mu}_{~\nu\rho\sigma}\del_\mu C_0
=
\mp\frac{e^{\phi}}{24}
 \bar\epsilon^{\mu}_{~\nu\rho\sigma}\del_\mu C_0\ .
\end{eqnarray}
Then, the relations
\eqref{map_first}-\eqref{map_last} imply (\ref{delac}) and (\ref{md}),
which is equivalent to the condition $e^{IJK}=0$ that solves
one of the SUSY condition (\ref{SUSYcond1}) as discussed in the previous section.

In the following section, we are going to check these identifications by
explicitly inserting some particular backgrounds in the effective action
for the D3-branes and comparing with the supersymmetric deformations of
the ${\cal N}=4$ SYM reviewed in Section \ref{sec:SYM}.

\section{Examples}
\label{sec:examples}

\subsection{$(p,q)$ 5-branes and Gaiotto-Witten solution}

In this subsection, we consider D3-branes embedded in a background with
$(p,q)$ 5-branes.\footnote{Here, $p$ and $q$ are relativley prime integers
and a $(p,q)$ 5-brane is a bound state of $p$ NS5-brane and $q$ D5-brane.}
The brane configuration is summarized as
\begin{eqnarray}
\begin{tabular}{c|ccccccccccccc}
&0&1&2&3&4&5&6&7&8&9\\
\hline
(probe) D3&o&o&o&o\\
$(p,q)\,$5&o&o&o&&o&o&o&&&
\end{tabular}
\label{D3D5}
\end{eqnarray}
The effective action on the D3-brane world-volume
can be written down by using \eqref{SboseD3} and \eqref{SfermiD3}.
As we will soon see, because the $(p,q)$ 5-branes are not extended along
the $x^3$-direction, the gauge coupling and the theta parameter of the
D3-brane action depend on the coordinate $x^3$. This brane configuration
is related to the supersymmetric Janus configurations considered in
\cite{Gaiotto:2008sd}.
We will show that the action obtained by using
 \eqref{SboseD3} and \eqref{SfermiD3} is indeed consistent with
that obtained in \cite{Gaiotto:2008sd}, which provides
a consistency check of our results in section \ref{sec:D3-curved}.

In this subsection, the letters for the indices are chosen as
 $\alpha,\beta=0,1,2$; $a,b,c=4,5,6$ and $p,q,r=7,8,9$.
Let us consider $n$ $(p,q)$ 5-branes placed at $x^3=0$, $x^p=x_0^p$ ($p=7,8,9$).
The supergravity solution corresponding to the $(p,q)$ 5-branes
can be obtained by applying $SL(2,\Z)$ duality to the D5-brane solution.
Its explicit form is
\footnote{See,{\it e.g.},
\cite{Bergshoeff:1995sq,Lu:1998vh,Llatas:1999zr}.}
\begin{align}
ds_E^2 &=
h(r)^{-\f{1}{4}}
(\eta_{\alpha\beta}dx^\alpha dx^\beta+\delta_{ab}dx^a dx^b)
+h(r)^{\f{3}{4}} ((dx^3)^2+\delta_{pq}dx^pdx^q)\ ,
\label{5metric}
\\
e^{-\phi}&=
\frac{\rho}{p^2g_s^{-1} h(r)^{1/2}+(q+p\chi_0)^2g_s h(r)^{-1/2}}
\ ,~~~
C_0=
\frac{pq (1-h(r))+\rho\chi_0g_s}{p^2g_s^{-1} h(r)+(q+p\chi_0)^2g_s}\ ,
\label{dilaton-C0}
\\
H_3&=2np\, l_s^2\epsilon_3\ ,~~~F_3=2nq\,g_s l_s^2\epsilon_3\ ,
\label{H3F3-1}
\end{align}
where $ds_E^2$ denotes the line element in the Einstein frame,
$\chi_0$ is a constant,
\begin{eqnarray}
h(r)\equiv 1+\frac{n\sqrt{\rho g_s}\,l_s^2}{r^2}\ ,~~~
\rho\equiv p^2g_s^{-1}+(q+p\chi_0)^2g_s\ ,~~~
r^2\equiv (x^3)^2+\sum_{p=7,8,9}(x^p-x_0^p)^2\ .
\end{eqnarray}
$\epsilon_3$ in (\ref{H3F3-1}) is the volume form of the unit $S^3$
embedded in the $\R^4$ parametrized by $x^{3,7,8,9}$
with its center at the position of the $(p,q)$ 5-brane.
$\epsilon_3$ can be written explicitly as
\begin{eqnarray}
 \epsilon_3= \sin^2\theta\sin\phi_1 d\theta\wedge d\phi_1\wedge d\phi_2\ ,
\end{eqnarray}
where $(\theta,\phi_1,\phi_2)$ are the coordinates on
the unit $S^3$ with $0\le\theta\le\pi$, $0\le\phi_1\le\pi$ and
$0\le\phi_2\le 2\pi$, related to $x^{3,7,8,9}$ as
\beal{
x^3 &= r \cos \theta\ , \nN
x^7-x_0^7 &=r \sin \theta \cos \phi_1\ ,\nN
x^8-x_0^8 &= r\sin \theta \sin \phi_1 \cos \phi_2\ ,\nN
x^9-x_0^9 &= r\sin \theta \sin \phi_1 \sin \phi_2\ .
\label{polar}
}

Note that $H_3$ and $F_3$ in (\ref{H3F3-1}) can be written as
\begin{eqnarray}
 H_{p'q'r'}
=\frac{p}{\sqrt{\rho g_s}}\,\varepsilon_{p'q'r'}^{~~~~~s'}\del_{s'}h(r)\ ,
~~~
 F_{p'q'r'}
=q\sqrt{\frac{g_s}{\rho}}\,\varepsilon_{p'q'r'}^{~~~~~s'}\del_{s'}h(r)\ ,
\label{H3F3-2}
\end{eqnarray}
where $p',q',r',s'=3,7,8,9$ and $\varepsilon_{p'q'r'}^{~~~~~s'}
=\varepsilon_{p'q'r't'}\delta^{t's'}$ is the
Levi-Civita symbol for the flat $\R^4$ parametrized
by $x^{3,7,8,9}$ with $\varepsilon_{3789}=+1$.\footnote{
One can easily recover the expressions of $H_3$ and $F_3$
in (\ref{H3F3-1}) from (\ref{H3F3-2}) by using the polar coordinates (\ref{polar})
with metric of $\R^4$: $ds^2=dr^2+r^2(d\theta^2+\sin^2\!\theta
d\phi_1^2+\sin^2\!\theta\sin^2\!\phi_1 d\phi_2^2)$ and
$\varepsilon_{r\theta\phi_1\phi_2}=r^3\sin^2\!\theta\sin\phi_1$.
}
In the the expressions (\ref{5metric}), (\ref{dilaton-C0}) and
(\ref{H3F3-2}), the function $h(r)$ can be replaced with an arbitrary
positive harmonic function on $\R^4$, which corresponds to a
supergravity solution describing parallel multiple $(p,q)$ 5-branes
distributed in $\R^4$.

The dilaton and R-R 0-from combined into a complex scalar field
$\tau\equiv g_s^{-1}(C_0+ie^{-\phi})$ can be written as
\begin{eqnarray}
\tau=\tau_0+4\pi D\,e^{i2\psi}\ ,
\label{5tau}
\end{eqnarray}
where
\begin{eqnarray}
\tau_0\equiv -\frac{q}{p}+4\pi D\ ,~~~
4\pi D\equiv \frac{\rho}{2p(q+p\chi_0)g_s}\ ,
\end{eqnarray}
are real constants and $\psi(r)$ is a real function satisfying
\begin{eqnarray}
 \tan\psi(r)=\frac{p\,h(r)^{1/2}}{(q+p\chi_0)g_s}\ .
\label{psi(r)}
\end{eqnarray}
Here, we have assumed $p,q,\chi_0$ are all positive and
$0<\psi<\frac{\pi}{2}$.
The complex scalar field (\ref{5tau}) evaluated at $x^i=0$
corresponds to the complex coupling (\ref{tau}).
In fact, the expression in (\ref{5tau}) agrees with 
the complex coupling obtained in \cite{Gaiotto:2008sd}. (See (\ref{tauGW}).)
Note here that $\psi|_{x^i=0}$ can be chosen to be
a generic real function of $x^3$, because
 as mentioned above, $h(r)$ in (\ref{psi(r)}) can be replaced with an
arbitrary positive harmonic function on $\R^4$ transverse to the
$(p,q)$ 5-branes.

The metric in the string frame is given by
\beal{ 
ds_{\rm string}^2 &= e^{\half\phi} ds_E^2 
\nN
& = \f{p}{\sqrt{\rho g_s}\,\sin\psi } \bigg[
(\eta_{\alpha\beta}dx^\alpha dx^\beta+\delta_{ab}dx^a dx^b)
+ \left(\f{\rho}{8\pi Dp^2}\tan\psi \right)^2
\left((dx^3)^2+\delta_{pq}dx^pdx^q\right) \bigg]\ .
\label{5metric-2}
}

It is easy to show that both $V_{\rm DBI}$
and $V_{\rm CS}$ in the potential (\ref{VPhi})
are flat (i.e. $\Phi^i$ independent),
because $F_5=0$, $H_{i\mu\nu}=F_{i\mu\nu}=0$, $B_{\mu\nu}=0$
and $L_{\rm DBI}$ defined in (\ref{LDBI}) is a constant:
\begin{eqnarray}
 L_{\rm DBI}=e^{-\phi}\sqrt{-\det(g_{\mu\nu})}=1\ .
\end{eqnarray}
Next, consider $N$ probe D3-branes placed at $x^i=0$
 ($i=4,\dots,9$) in this background.
The metric (\ref{5metric-2}) evaluated at $x^i=0$ is written as
\beal{ 
\left.ds_{\rm string}^2\right|_{x^i=0}
=e^{\xi}\left[\eta_{\mu\nu}dy^\mu dy^\nu
+\delta_{ab}dx^adx^b+e^{2\eta}\delta_{pq}dx^pdx^q
\right]\ ,
\label{5metric-3}
}
where we have defined
\begin{eqnarray}
e^{\xi}\equiv
\left.\f{p}{\sqrt{\rho g_s}\,\sin \psi(r) }\right|_{x^i=0}\ ,~~~
e^{\eta}\equiv\left. h(r)^{1/2}\right|_{x^i=0}=
\left.\f{\rho}{8\pi Dp^2}\tan \psi(r)\right|_{x^i=0}\ ,
\end{eqnarray}
and introduced new coordinates $y^\mu$ ($\mu=0,1,2,3$) satisfying
\begin{eqnarray}
y^0=x^0\ ,~~y^1=x^1\ ,~~y^2=x^2\ ,~~
dy^3=e^\eta dx^3\ .
\end{eqnarray}
Then, using the coordinates $y^\mu$,
the bosonic part of the effective action (\ref{SboseD3}) becomes
\begin{align}
S_{\rm D3}^{\rm boson}
=&
\frac{T_3\lambda^2}{2}\int d^4y\,
e^{-\phi}\tr\bigg(
-\frac{1}{2}\eta^{\mu\nu}\eta^{\rho\sigma}F_{\mu\rho}F_{\nu\sigma}
\pm\frac{e^\phi C_0}{4}
\epsilon^{\mu\nu\rho\sigma}F_{\mu\nu}F_{\rho\sigma}
\nn\\
&\quad
+e^{2\xi}\eta^{\mu\nu}\delta_{ab}D_\mu\Phi^aD_\nu\Phi^b
+e^{2(\xi+\eta)}\eta^{\mu\nu}\delta_{pq}D_\mu\Phi^pD_\nu\Phi^q
\nn\\
&+\half
e^{4\xi}\delta_{ab}\delta_{a'b'}[\Phi^a,\Phi^b][\Phi^{a'},\Phi^{b'}]
+\half
 e^{4(\xi+\eta)}\delta_{pq}\delta_{p'q'}[\Phi^p,\Phi^q][\Phi^{p'},\Phi^{q'}]
\nn\\
&
+e^{4\xi+2\eta}\delta_{ab}\delta_{pq}[\Phi^a,\Phi^b][\Phi^{p},\Phi^{q}]
\mp\frac{i}{3}e^{2\xi+\phi}(G^R_\pm)_{ijk}\Phi^i[\Phi^j,\Phi^k]
\bigg)\ .
\label{5Sboson}
\end{align}

In order to compare with the action (\ref{action1}), it is convenient to
rescale the scalar fields as
\begin{eqnarray}
 A_a\equiv e^{\xi}\Phi^a\ ,~~~A_p\equiv e^{\xi+\eta}\Phi^p\ .
\end{eqnarray}
Then, the kinetic term of the scalar fields can be rewritten as
\begin{align}
&
e^{-\phi}{\rm tr}\left(
e^{2\xi}\eta^{\mu\nu}\delta_{ab}D_\mu\Phi^aD_\nu\Phi^b+
e^{2(\xi+\eta)}\eta^{\mu\nu}\delta_{pq}D_\mu\Phi^pD_\nu\Phi^q
\right)
\nN
=&\
e^{-\phi}{\rm tr}\left(
\eta^{\mu\nu}\delta^{AB}D_\mu A_AD_\nu A_B
+\half m^{AB}A_AA_B
\right)+(\mbox{total derivative})\ ,
\end{align}
where $A,B=4,\dots,9$ and the non-zero components of $m^{AB}$
are
\begin{align}
m^{ab}&
=2\left(\xi'^2+\xi''-\phi'\xi'\right)\delta^{ab}
=2\left(\psi'^2-(\psi'\cot\psi)'\right)\delta^{ab}
\ ,
\nn\\
 m^{pq}&
=2\left((\xi'+\eta')^2+\xi''+\eta''-\phi'(\xi'+\eta')\right)\delta^{pq}
=2\left(\psi'^2+(\psi'\tan\psi)'\right)\delta^{pq}
\ .
\end{align}
Here, the prime denotes the derivative with respect to $y^3$,{\it e.g.},
$\xi'=\del_{y^3}\xi$.
These expressions precisely agree with (\ref{mGW})

Note that the non-zero components of $(G_\pm^{R})_{ijk}$ are
\begin{align}
&(G_\pm^{R})_{456}
=-e^{-3\eta}(F_{789}+C_0H_{789})
=2\sqrt{\frac{g_s}{\rho}}(q+g_s^{-1}C_0p)\eta'\ ,
\\
&(G_\pm^{R})_{789}
=\mp e^{-\phi}H_{789}
=\pm e^{-\phi}\frac{2p}{\sqrt{\rho g_s}}e^{3\eta}\eta'\ ,
\end{align}
and one can show the following relations:
\begin{eqnarray}
e^{\phi-\xi}(G_\pm^{\rm R})_{456}
=\frac{2\psi'}{\sin\psi}\ ,~~~
e^{\phi-\xi-3\eta}(G_\pm^{\rm R})_{789}
=\pm \frac{2\psi'}{\cos\psi}\ .
\label{Grel}
\end{eqnarray}

Using these, the last term of (\ref{5Sboson}) can be written as
\begin{align}
\mp\frac{i}{3}e^{2\xi+\phi}
(G_\pm^{\rm R})_{ijk}\Phi^i[\Phi^j,\Phi^k]
&=\mp 2i
\left(e^{\phi-\xi}(G_\pm^{\rm R})_{456}A_4[A_5,A_6]
+e^{\phi-\xi-3\eta}(G_\pm^{\rm R})_{789} A_7[A_8,A_9]
\right)
\nN
&=
4i\left(\mp\frac{\psi'}{\sin\psi}A_4[A_5,A_6]
- \frac{\psi'}{\cos\psi}A_7[A_8,A_9]
\right)\ .
\end{align}
These terms (with the upper sign) agree with (\ref{dGW}).
(The lower sign is obtained,{\it e.g}., by
a transformation $(x^1,x^4)\ra (-x^1,-x^4)$.)

In summary, the bosonic part is written as
\begin{align}
S_{\rm D3}^{\rm boson}
=&
\frac{T_3\lambda^2}{2}\int d^4y\,
e^{-\phi}\tr\bigg(
-\frac{1}{2}\eta^{II'}\eta^{JJ'}F_{IJ}F_{I'J'}
\pm\frac{e^\phi C_0}{4}
\epsilon^{\mu\nu\rho\sigma}F_{\mu\nu}F_{\rho\sigma}
\nn\\
&
+4i\left(
\mp\frac{\psi'}{\sin\psi}A_4[A_5,A_6]
- \frac{\psi'}{\cos\psi}A_7[A_8,A_9]
\right)
\nn\\
&
-(\psi'^2-(\psi'\cot\psi)'))\delta^{ab}A_aA_b
-(\psi'^2+(\psi'\tan\psi)'))\delta^{pq}A_pA_q
\bigg)
\end{align}
with $\phi$ and $C_0$ given by (\ref{5tau}).

Let us next consider to the fermionic part.
The action (\ref{SfermiD3}) in the background
(\ref{5metric})--(\ref{H3F3-1}) is
\begin{align}
 S_{\rm D3}^{\rm fermi}
=&\
\frac{T_3\lambda^2}{2}\int d^4 y\, e^{2\xi}\,e^{-\phi}\tr \left\{
i(\ol\Psi\Gamma^\mu D_\mu\Psi
+\ol\Psi\Gamma_k i[\Phi^k,\Psi])
-i\ol\Psi M_\pm \Psi \right\}
\end{align}
with
\begin{eqnarray}
M_\pm \equiv \mp \frac{e^{\phi}}{4}\left(
(*_4 F_1)_{012}\Gamma^{012}
+(G_\pm^R)_{456}\Gamma^{456}
+(G_\pm^R)_{789}\Gamma^{789}
\right)
\ .
\end{eqnarray}
Rescaling $\Psi$, $M_\pm$ and the gamma matrices as
\begin{eqnarray}
&&\wh\Psi\equiv e^{\frac{3}{4}\xi}\Psi\ ,~~
\wh M_\pm\equiv e^{\xi/2}M_\pm\ ,
\nn\\
&&\wh\Gamma^\mu\equiv e^{\xi/2}\Gamma^\mu\ ,~~
\wh\Gamma_a\equiv
\wh\Gamma^a\equiv e^{\xi/2+\eta}\Gamma^a
=e^{-(\xi/2+\eta)}\Gamma_a\ ,~~
\wh\Gamma_p\equiv\wh\Gamma^p\equiv
e^{\xi/2}\Gamma^p=e^{-\xi/2}\Gamma_p\ ,
\end{eqnarray}
we obtain
\begin{align}
S_{\rm D3}^{\rm fermi}
=&\
\frac{T_3\lambda^2}{2}\int d^4 y\,e^{-\phi}\tr \left\{
i(\ol{\wh\Psi}\wh\Gamma^\mu D_\mu\wh\Psi
+\ol{\wh\Psi}\wh\Gamma^A i[A_A,\Psi])
-i\ol{\wh\Psi}\wh M_\pm\wh\Psi \right\}\ .
\end{align}
Here, the rescaled gamma matrices $\wh\Gamma^I$ ($I=0,1,\dots,9$)
satisfy the anti-commutation relations with the flat metric
$\{\wh\Gamma^I,\wh\Gamma^J\}=\eta^{IJ}$, and we have used
$\ol\Psi\Gamma^\mu(\del_\mu\xi)\Psi=0$, which follows because
$\Gamma^0\Gamma^\mu$ is a symmetric matrix.

Again, using (\ref{Grel}), we obtain
\begin{align}
\wh M_\pm
&=\mp \frac{1}{4}\left(
e^{\phi}\del_3 C_0\wh\Gamma^{012}
+e^{\phi-\xi}(G_\pm^R)_{456}\wh\Gamma^{456}
+e^{\phi-\xi-3\eta}(G_\pm^R)_{789}\wh\Gamma^{789}
\right)
\nN
&= \frac{1}{2}\left(\pm\psi'\,\wh\Gamma^{012}
\mp\frac{\psi'}{\sin\psi}\,\wh\Gamma^{456}
- \frac{\psi'}{\cos\psi}\,\wh\Gamma^{789}
\right)\ ,
\end{align}
which reproduces (\ref{MfermiGW}).
(Again, the lower sign is obtained by the transformation
$(x^1,x^4)\ra (-x^1,-x^4)$.)

\subsection{Backgrounds with D3-branes}

Let us next consider probe D3-branes extended along
$x^{0,1,2,3}$ directions in a background corresponding
to $n$ D3-branes extended along $x^{0,1,4,5}$ directions:
\begin{eqnarray}
\begin{tabular}{c|ccccccccccccc}
&0&1&2&3&4&5&6&7&8&9\\
\hline
(probe) D3&o&o&o&o\\
D3&o&o&&&o&o&&&&
\end{tabular}
\label{D3D3}
\end{eqnarray}
In this subsection, we use the letters for the
indices as $\alpha,\beta=0,1$; $m,n=2,3$;
$a,b=4,5$ and $p,q=6,7,8,9$.

The supergravity solution corresponding to $n$ D3-branes
placed at $x^m=0$ and $x^p=x_0^p$,
in the string frame, is
\beal{
&e^{\phi}=1\ ,~~~C_0 = \text{constant}\ , \\
&ds^2_{\rm string} =
 h(r)^{-\half}(\eta_{\alpha\beta}dx^\alpha dx^\beta
+\delta_{ab}dx^a dx^b)
+ h(r)^\half (\delta_{mn}dx^m dx^n+\delta_{pq}dx^p dx^q)\ ,\\
&F_5 =f_5+*f_5\ ,~~~
f_5\equiv d h(r)^{-1}\wedge dx^0 \wedge dx^1 \wedge dx^4 \wedge dx^5 ,
}
where $g_s$ is a constant and $h(r)$ is given as
\beal{
h(r)\equiv 1+\f{Q_3}{r^4}\ ,~~~ Q_3\equiv 4\pi g_s n l_s^4\ ,
~~~r^2\equiv
\sum_{m=2,3} (x^m)^2+\sum_{p=6}^9(x^p-x_0^p)^2\ .
\label{h(r)-D3}
}
As in the previous subsection, the function $h(r)$ can be
replaced with an arbitrary positive harmonic function
on the $\R^6$ parametrized by $x^{2,3,6,7,8,9}$.

The metric evaluated at the position of the probe D3-branes, {\it i.e.}
 $x^i=0$ ($i=4,\dots,9$), is
\begin{eqnarray}
\left. ds^2_{\rm string}\right|_{x^i=0}
=e^{-\half\varphi}(\bar g_{\mu\nu}dx^\mu dx^\nu
+\delta_{ab}dx^adx^b+e^\varphi\delta_{pq}dx^pdx^q 
)\ ,
\label{metric2-4}
\end{eqnarray}
where we have defined
\begin{eqnarray}
e^{\varphi(x^m)}\equiv h(r)|_{x^i=0}\ ,
\end{eqnarray}
and
\begin{eqnarray}
\bar g_{\mu\nu}dx^\mu dx^\nu
\equiv \eta_{\alpha\beta}dx^\alpha dx^\beta+e^\varphi
\delta_{mn}dx^m dx^n\ .
\end{eqnarray}
This metric (\ref{metric2-4}) has $ISO(1,1)\times SO(2)\times SO(4)$
isometry, where $ISO(1,1)$ is the Poincar\'e symmetry acting
on $x^{0,1}$, $SO(2)$ and $SO(4)$ are rotational symmetry
acting on $x^{4,5}$ and $x^{6,7,8,9}$, respectively.\footnote{
If we use $h(r)$ in (\ref{h(r)-D3}), the metric
also has a rotational symmetry on the $x^{2,3}$ plane.
However, as mentioned above, $h(r)$ can be replace with
an arbitrary positive harmonic function on $\R^6$, in
which case the rotational symmetry on $x^{2,3}$
is broken in general.
}
The supersymmetry condition for the deformed ${\cal N}=4$
SYM (\ref{action1}) with this symmetry is analyzed in
section 4.1 of \cite{Choi:2017kxf}. Since our brane configuration
(\ref{D3D3}) preserves part of the supersymmetry,
the action (\ref{SboseD3}) and (\ref{SfermiD3}) for this background
should reproduce one of the general solutions obtained there.

It is again easy to see that both $V_{\rm DBI}$ and $V_{\rm CS}$
in the potential (\ref{VPhi}) are flat, because
$F_3=H_3=0$, $F_{\mu\nu\rho\sigma i}=0$ and $L_{\rm DBI}$
defined in (\ref{LDBI}) is a constant.
Then, the bosonic part (\ref{SboseD3}) becomes
\begin{align}
S_{\rm D3}^{\rm boson}
=&\
\frac{T_3\lambda^2}{2}\int d^4 x
\sqrt{-\bar g}
\tr\bigg\{
-\frac{1}{2}
 \bar g^{\mu\nu}\bar g^{\rho\sigma}F_{\mu\rho}F_{\nu\sigma}
\pm\frac{C_0}{4}
\bar\epsilon^{\mu\nu\rho\sigma}F_{\mu\nu}F_{\rho\sigma}
\nn\\
&
-\bar g^{\mu\nu}
(e^{-\varphi}\delta_{ab}D_\mu\Phi^a D_\nu\Phi^b
+\delta_{pq}D_\mu\Phi^p D_\nu\Phi^q)
\nn\\
&+\half
e^{-2\varphi}\delta_{ab}\delta_{a'b'}[\Phi^a,\Phi^b][\Phi^{a'},\Phi^{b'}]
+\half
\delta_{pq}\delta_{p'q'}[\Phi^p,\Phi^q][\Phi^{p'},\Phi^{q'}]
\nn\\
&
+e^{-\varphi}\delta_{ab}\delta_{pq}[\Phi^a,\Phi^b][\Phi^{p},\Phi^{q}]
\mp e^{-\varphi}(*_4\wt F_5)^\mu_{~ij}\Phi^iD_\mu\Phi^j
\bigg\}\ ,
\label{3bosonic}
\end{align}
where $\bar\epsilon^{\mu\nu\rho\sigma}$ is the Levi-Civita symbol
with $\bar\epsilon^{0123}\equiv 1/\sqrt{-\bar g}$.
In order to compare with the results in \cite{Choi:2017kxf},
we redefine the scalar fields as
\begin{eqnarray}
 A_a\equiv e^{-\varphi/2}\Phi^a\ ,~~~
 A_p\equiv \Phi^p\ .
\end{eqnarray}
Then, the kinetic terms for the scalar fields become
\begin{eqnarray}
&&
\sqrt{-\bar g}\,{\rm tr}\left(
e^{-\varphi}\bar g^{\mu\nu}\delta_{ab}D_\mu\Phi^aD_\nu\Phi^b+
\bar g^{\mu\nu}\delta_{pq}D_\mu\Phi^pD_\nu\Phi^q
\right)
\nN
&=&
\sqrt{-\bar g}\,{\rm tr}\left(
\bar g^{\mu\nu}\delta^{AB}D_\mu A_AD_\nu A_B
+\half m^{AB}A_AA_B
\right)+(\mbox{total derivative})\ ,
\end{eqnarray}
where
\begin{align}
m^{ab}=\bar g^{mn}\left(\half\del_m\varphi\del_n\varphi
-\del_m\del_n\varphi\right)\delta^{ab}\ ,~~~
 m^{pq}=0\ .
\label{mab}
\end{align}

The non-zero components of $(*_4\wt F_5)^\mu_{~ij}$ are
\begin{eqnarray}
(*_4\wt F_5)^n_{~45}
= -\bar\epsilon^{mn}\del_m\varphi\ ,
\end{eqnarray}
where $\bar\epsilon^{23}=-\bar\epsilon^{32}
=\sqrt{\bar g^{22}\bar g^{33}}=e^{-\varphi}$.
The last term in (\ref{3bosonic}) becomes
\begin{eqnarray}
e^{-\varphi}(*_4\wt F_5)^\mu_{~ij}\Phi^iD_\mu\Phi^j
=
-\bar\epsilon^{mn}\del_m\varphi\,\varepsilon^{ab}A_aD_nA_b\ ,
\end{eqnarray}
where
$\varepsilon^{45}=-\varepsilon^{54}=1$.
This gives
\begin{eqnarray}
 d^{nab}=\pm\half\bar\epsilon^{mn}\del_m\varphi\,\varepsilon^{ab}\ ,
\label{dnab}
\end{eqnarray}
in (\ref{action1}).

Collecting all these results, (\ref{3bosonic}) becomes
\begin{align}
S_{\rm D3}^{\rm boson}
=&
\frac{T_3\lambda^2}{2}\int d^4 x
\sqrt{-\bar g}
\tr\bigg\{
-\frac{1}{2}
 \bar g^{II'}\bar g^{JJ'}F_{IJ}F_{I'J'}
\pm\frac{C_0}{4}
\bar\epsilon^{\mu\nu\rho\sigma}F_{\mu\nu}F_{\rho\sigma}
\nn\\
&
-\frac{1}{2}
\bar g^{mn}\left(\half\del_m\varphi\del_n\varphi
-\del_m\del_n\varphi\right)\delta^{ab}A_aA_b
\pm\bar\epsilon^{mn}\del_m\varphi\,\varepsilon^{ab}
A_a D_n A_b
\bigg\}\ .
\label{3bosonic-2}
\end{align}

The fermionic part (\ref{SfermiD3}) for this configuration is
\begin{align}
 S_{\rm D3}^{\rm fermi}
=&\
\frac{T_3\lambda^2}{2}\int d^4 x\, \sqrt{-\bar g}\, e^{-\varphi}\tr \left\{
i(\ol\Psi\Gamma^\mu D_\mu\Psi
+\ol\Psi\Gamma_k i[\Phi^k,\Psi])
-i\ol\Psi M_\pm \Psi \right\}
\label{3fermi}
\end{align}
with
\begin{eqnarray}
M_\pm =\pm\frac{1}{4}(*_4\wt F_5)^{n}_{~ 45}\Gamma^{m 45}g_{mn}\ .
\end{eqnarray}
As in the previous subsection, we
rescale $\Psi$, $M_\pm$ and the gamma matrices by
\begin{eqnarray}
&&\wh\Psi\equiv e^{-\frac{3}{8}\varphi}\Psi\ ,~~~
\wh M_\pm\equiv M_\pm e^{-\frac{1}{4}\varphi}\ ,
\nn\\
&&\wh\Gamma^\mu\equiv e^{-\frac{1}{4}\varphi}\Gamma^\mu\ ,~~~
\wh\Gamma_a\equiv\wh\Gamma^a\equiv e^{-\frac{1}{4}\varphi}\Gamma^a
= e^{\frac{1}{4}\varphi}\Gamma_a\ ,~~~
\wh\Gamma_p\equiv\wh\Gamma^p\equiv e^{\frac{1}{4}\varphi}\Gamma^p
= e^{-\frac{1}{4}\varphi}\Gamma_p\ .
\end{eqnarray}
The rescaled gamma matrices satisfy
\begin{eqnarray}
 \{\wh\Gamma^\mu,\wh\Gamma^\nu\}=2\bar g^{\mu\nu}\ ,~~~
 \{\wh\Gamma^a,\wh\Gamma^b\}=2\delta^{ab}\ ,~~~
 \{\wh\Gamma^p,\wh\Gamma^q\}=2\delta^{pq}\ .
\end{eqnarray}
Then, we obtain
\begin{eqnarray}
 S_{\rm D3}^{\rm fermi}
=
\frac{T_3\lambda^2}{2}\int d^4 x\, \sqrt{-\bar g}\, \tr \left\{
i\left(\ol{\wh\Psi}\wh\Gamma^\mu D_\mu\wh\Psi
+\ol{\wh\Psi}\wh\Gamma^A i[A_A,\wh\Psi]\right)
-i\ol{\wh\Psi}\wh M_\pm \wh\Psi \right\}
\label{3fermi-2}
\end{eqnarray}
with
\begin{eqnarray}
\wh M_\pm=\pm\frac{1}{4}(*\wt F_5)^n_{~45}\wh\Gamma^{m45}\bar g_{nm}
=\mp \frac{1}{4}\bar\epsilon^{n}_{~m}\del_{n}\varphi\wh\Gamma^{m45}\ ,
\end{eqnarray}
where $\bar\epsilon^n_{~m}\equiv\bar\epsilon^{nn'}\bar g_{n'm}$.
This gives
\begin{eqnarray}
\beta_m=\mp\frac{1}{4}\bar\epsilon^{n}_{~m}\del_{n}\varphi\ .
\label{betam}
\end{eqnarray}

The results (\ref{mab}), (\ref{dnab}) and (\ref{betam}) agree
with (\ref{mab2-4}), (\ref{dnab2-4}) and (\ref{beta2-4}), respectively,
for the case with $\tau=\mbox{constant}$, $\Lambda=\mbox{constant}$
and $s=\mp$.\footnote{See the case (C3) with
$\beta_m=\mp\frac{1}{4}\bar\epsilon^{n}_{~m}\del_{n}\varphi$
in section 4.1.2 of \cite{Choi:2017kxf}.
}

\section{Conclusions and outlook}
\label{sec:conclusions}

In this work, we have complemented the study of deformations of $\cN=4$
SYM with varying couplings that we initiated in \cite{Choi:2017kxf} by
showing that some of these gauge theories can be realized on the probe
D3-branes in curved backgrounds with fluxes. 
In particular, we obtained the effective action on the D3-branes
for general backgrounds satisfying (\ref{assumptions})
and gave an explicit map between the couplings
in the deformed ${\cal N}=4$ SYM and the fluxes of the curved
background on which the D3-branes are embedded.

As a check,
we explicitly showed that the effective action on the D3-branes
in a background with $(p,q)$ 5-branes (see \eqref{D3D5})
reproduces that of the supersymmetric Janus configuration found
in \cite{Gaiotto:2008sd}. We also studied 
D3-branes in a background with another stack of
D3-branes intersecting with them (see \eqref{D3D3}) and 
found that the action agrees with one of the solutions
of SUSY conditions with $ISO(1,1)\times SO(2)\times SO(4)$ symmetry
found in \cite{Choi:2017kxf}.

On the other hand, in \cite{Choi:2017kxf}, we found a lot of
solutions of SUSY conditions, for which the realization
in string theory is not known. 
Our results in (\ref{map_first})--(\ref{map_last}) suggest
that it is possible to extract some information of supergravity
fields from the couplings in the deformed $\cN=4$ SYM.
Indeed, it is now easy to know which fluxes have non-trivial profiles
for the brane configuration that realizes the deformed $\cN=4$ SYM.
For example,
for the cases with $ISO(1,1)\times SO(3)\times SO(3)$ symmetry,
solutions with non-trivial $m_{012}$, $m_{013}$,
$m_{456}$ and $m_{789}$ are found in \cite{Choi:2017kxf}.
Such configurations, assuming that they can be realized in string theory,
should have non-trivial $(*_4F_1)_{012}$,
 $(*_4F_1)_{013}$, $(G_\pm^R)_{456}$ and $(G_\pm^R)_{789}$ fluxes.
Despite we have not shown this explicitly, this fact suggests
that such a configuration corresponds to D3-branes in a background with
$(p,q)5$- and $[p',q']7$-branes: 
\begin{eqnarray}
\begin{tabular}{c|ccccccccccccc}
&0&1&2&3&4&5&6&7&8&9\\
\hline
(probe) D3&o&o&o&o
\\
$(p,q)$ 5 &o&o&o&&o&o&o&&&
\\
$[p',q']$ 7 &o&o&&&o&o&o&o&o&o
\end{tabular}
\label{D3D5D7}
\end{eqnarray}
It would be interesting to see this more explicitly.

Finally, we want to stress that we didn't use
the equations of motion for the supergravity fields
in our analysis in Section \ref{sec:D3-curved}.
That is to say, some additional constraints are imposed on
the couplings from the supergravity equations of motion. In
this respect, some works has been done in
\cite{Camara:2003ku,Camara:2004jj}, where it has been shown that the
couplings have to satisfy some algebraic equations obtained from
the supergravity equations of motion.
Furthermore, if we require SUSY, the background as well as the
D3-brane configurations should satisfy BPS conditions.
It would be interesting to see whether such conditions agree with
the SUSY conditions found in \cite{Choi:2017kxf}.
Actually, there is a logical possibility that
the deformed $\cN=4$ SYM action \eqref{action1}
could have some additional SUSY solutions which are not necessarily
related to backgrounds satisfying the equations of motion in supergravity.
It would be important to study the correspondence
in more detail and clarify this issue.

\section*{Acknowledgements}

We thank Adi Armoni, Ki-Myeong Lee,
Takahiro Nishinaka, Yuta Sekiguchi and Shuichi Yokoyama
for discussion.
We also appreciate useful discussions during the workshops ``Geometry,
Duality and Strings'' Ref. YITP-X-16-11, ``Strings and Fields''
Ref. YITP-W-17-08 and the KIAS-YITP Joint Workshop 2017 ``Strings,
Gravity and Cosmology'' YITP-W-17-12, held at the Yukawa Institute for
Theoretical Physics, Kyoto University.
J.J.F.-M. gratefully acknowledges the support of JSPS (Postdoctoral
Fellowship) and the Fundaci\'on S\'eneca/Universidad de Murcia (Programa
Saavedra Fajardo).
The work of S.S. was supported by JSPS KAKENHI (Grant-in-Aid for
Scientific Research (C)) Grant Number JP16K05324.
The work of J.J.F.-M. and S.S. was also supported by JSPS KAKENHI
(Grant-in-Aid for JSPS Fellows) Grant Number JP16F16741.

\appendix

\section{Conventions for supergravity fields}
\label{app:SUGRA}

We follow the conventions for the supergravity fields
used in \cite{Myers:1999ps}.
The bosonic part of the type IIB supergravity action in the string frame is
\begin{align}
 S_{\rm IIB}=&\ \frac{1}{2\kappa^2}
\int d^{10}x\sqrt{-g}\bigg\{e^{-2\phi}
\left(R+4|d\phi|^2-\half|H_3|^2\right)
-\half\left(|F_1|^2+|\wt F_3|^2+\half|\wt F_5|^2
\right)
\bigg\}
\nn\\
&+\frac{1}{4\kappa^2}\int\left(C_4+\half B_2\wedge C_2\right)
\wedge F_3\wedge H_3\ ,
\end{align}
where
\begin{eqnarray}
H_3=dB_2\ ,~~~
 F_n=dC_{n-1}\ ,~~~
\wt F_n=F_n+H_3\wedge C_{n-3}\ .
\label{H3Fn}
\end{eqnarray}
and   $|\omega_n|^2$ for an $n$-form $\omega_n$ is defined as
\begin{eqnarray}
|\omega_n|^2\equiv
\frac{1}{n!}\omega_{I_1\cdots I_n}\omega_{J_1\cdots J_n}g^{I_1J_1}
\cdots g^{I_nJ_n}\ .
\label{norm}
\end{eqnarray}

In our convention, the dilaton $\phi$ vanishes asymptotically
and $\kappa$ is related to the Newton's constant $G_N$, string length $l_s$
and string coupling $g_s$ as
\begin{eqnarray}
 2\kappa^2=16\pi G_N=(2\pi)^7l_s^8g_s^2\ .
\end{eqnarray}

In addition, we have to impose the self-duality condition
\begin{eqnarray}
 \wt F_5=*\wt F_5\ .
\label{F5selfdual}
\end{eqnarray}
Here, the Hodge star $*$ is defined by
\begin{eqnarray}
*(dx^{I_1}\wedge\cdots\wedge dx^{I_n})
=\frac{1}{(10-n)!}\epsilon^{I_1\cdots I_{10}}
g_{I_{n+1}J_{n+1}}\cdots g_{I_{10}J_{10}}
dx^{J_{n+1}}\wedge\cdots\wedge dx^{J_{10}}\ ,
\end{eqnarray}
where $\epsilon^{M_1\cdots M_{10}}$ is the 10-dimensional Levi-Civita symbol
with $\epsilon^{01 \cdots 9}=1/\sqrt{-g}$.

It is useful to define  $\wt F_{n}$ with $n>5$ by
\begin{eqnarray}
 \wt F_{n}\equiv (-1)^{\half n(n+1)+1}* \wt F_{10-n}\ .
\end{eqnarray}
Then, the equations of motion and the Bianchi identities 
for the R-R fields are written as
\begin{eqnarray}
 d\wt F_n+H_3\wedge \wt F_{n-2}=0\ ,~~~
(n=1,3,5,7,9)\ ,
\end{eqnarray}
which allows us to introduce $C_{n-1}$
satisfying (\ref{H3Fn}) for $n=1,3,5,7,9$.

$\phi$, $B_2$, $C_0$, $C_2$ and $C_4$
are related to those used
in \cite{Polchinski:1998rr}, denoted with superscript ``${\rm P}$'',
as
\begin{eqnarray}
e^{\phi^{\rm P}}=g_s e^\phi\ ,~~
B^{\rm P}_2=-B_2\ ,~~
C_0^{\rm P}=g_s^{-1}C_0\ ,~~
C_2^{\rm P}=g_s^{-1}C_2\ ,~~
C_4^{\rm P}=g_s^{-1}\left(C_4+\half B_2\wedge C_2\right)\ .
\end{eqnarray}

The metric in the Einstein frame is defined as
\begin{eqnarray}
 g^{\rm E}_{IJ}=e^{-\half\phi}g_{IJ}\ .
\end{eqnarray}
The action can be written as
\begin{eqnarray}
S_{\rm IIB}
&=&\frac{1}{2\kappa^2}\int d^{10}x\sqrt{-g^{\rm E}}\bigg\{
R^{\rm E}
-\half\left(|d\phi|_{\rm E}^2+e^{-\phi}|H_3|_{\rm E}^2+e^{2\phi}|F_1|_{\rm E}^2
+e^{\phi}|\wt F_3|_{\rm E}^2+\half|\wt F_5|_{\rm E}^2
\right)
\bigg\}
\nn\\
&&+\frac{1}{4\kappa^2}\int\left(C_4+\half B_2\wedge C_2\right)
\wedge F_3\wedge H_3
\nn\\
&=&\frac{1}{2\kappa^2}\int d^{10}x\sqrt{-g^{\rm E}}\bigg\{
R^{\rm E}
-\half\left(g_{\rm E}^{MN}\frac{\del_M\ol\tau\del_N\tau}{({\rm Im}\,\tau)^2}
+{\cal M}_{ij}F_3^i\cdot F_3^j
+\half|\wt F_5|_{\rm E}^2
\right)
\bigg\}
\nn\\
&&+\frac{\varepsilon_{ij}}{8\kappa^2}\int\left(C_4+
\half B_2\wedge C_2\right)
\wedge F_3^i\wedge F_3^j\ ,
\end{eqnarray}
where $|\omega_n|_{\rm E}^2$ is defined as in (\ref{norm}) with the
metric in the Einstein frame,
\begin{eqnarray}
\tau\equiv g_s^{-1}(C_0+ie^{-\phi})\ ,~~~
 F_3^1\equiv -g_s^{1/2}H_3\ ,~~~ F_3^2\equiv g_s^{-1/2}F_3\ ,
\end{eqnarray}
\begin{eqnarray}
 ({\cal M}_{ij})=\frac{1}{{\rm Im}\,\tau}
\mat{|\tau|^2,-{\rm Re}\,\tau,-{\rm Re}\,\tau,1}\ ,~~~
(\varepsilon_{ij})=\mat{,1,-1,}
\end{eqnarray}
and
\begin{eqnarray}
 F_3^i\cdot F_3^j\equiv
\frac{1}{3!}
F_{I_1I_2I_3}^i F_{J_1J_2J_3}^jg_{\rm E}^{I_1J_1}g_{\rm E}^{I_2J_2}g_{\rm E}^{I_3J_3}\ . 
\end{eqnarray}
This action is invariant under the $SL(2,\R)$ transformation:
\begin{eqnarray}
 \tau\ra\frac{a\tau+b}{c\tau+d}\ ,~~~
\left({F_3^2\atop F_3^1}\right)\ra
\mat{a,b,c,d}\left({F_3^2\atop F_3^1}\right)\ ,~~~
\mat{a,b,c,d}\in SL(2,\R)\ ,
\end{eqnarray}
with $\kappa$, $g_{MN}^{\rm E}$ and $C_4+\half B_2\wedge C_2$ kept fixed.

\section{Derivation of the D3-brane effective action}
\label{app:D3action}

In this appendix, we show the detailed derivation of the action
(\ref{SboseD3}) and (\ref{SfermiD3}).

\subsection{D$p$-branes in curved backgrounds (review)}
\label{Dpreview}

For convenience, we first review the effective action of
D$p$-branes embedded in general backgrounds in Appendix \ref{Dpreview},
following \cite{Myers:1999ps} and \cite{Martucci:2005rb} for bosonic
and fermionic parts, respectively.

\subsubsection{Bosonic part}

In this subsection, we review the bosonic part of
the effective action on D$p$-branes
embedded in a curved background with fluxes
following \cite{Myers:1999ps}.

The 10-dimensional space-time coordinates are denoted
as $x^I$ ($I=0,1,\dots,9$).
We choose the static gauge in which $x^\mu$ ($\mu=0,1,2,3$)
are identified as the coordinates on the D$p$-brane world-volume
and $x^i$ ($i=4,\dots,9$) parametrize the transverse directions.
The bosonic sector of the effective theory contains
a $U(N)$ gauge field $A_\mu$ ($\mu=0,\dots,p$),
$(9-p)$ scalar fields $\Phi^i$ ($i=p+1,\dots,9$),
which belong to the adjoint representation
of the gauge group $U(N)$.
The reference position of the D$p$-brane is chosen to be $x^i=0$
and small deviations from it is described by
the values of the scalar fields.

The effective action that describes the light open-string bosonic
fluctuations of a set of $N$ coincident D$p$-branes in type II string
theory consists of
\begin{align}
S_{{\rm D}p}^{\text{boson}}
=
S_{{\rm D}p}^{\rm DBI}+ S_{{\rm D}p}^{\rm CS}
\ ,
\label{bosonic-initial}
\end{align}
where the Dirac-Born-Infeld (DBI) and Chern-Simons (CS) terms
are given by
\begin{align}
S_{{\rm D}p}^{\rm DBI}
=&\
-T_{p}\int
 d^{p+1}x\,{\rm Str}\left\{e^{-\wh\phi}
\sqrt{-\det(M_{\mu\nu})\det(Q^i_{~j})}
\right\}
\ ,
\label{DBI-initial}
\\
S_{{\rm D}p}^{\rm CS}=&\
\mu_{p}\int
{\rm Str}\left\{{\rm P}
\left[e^{i\lambda \imath_\Phi\imath_\Phi}
\left(\sum_n\wh C_{n}\wedge e^{\wh B_2}\right)\right]
\wedge e^{\lambda F}
\right\}\ .
\label{CS-initial}
\end{align}
Here,
the parameters $T_p$, $\mu_p$ and $\lambda$ are given by
\begin{align}
T_p\equiv \frac{1}{(2\pi)^pl_s^{p+1}g_s}\ ,~~~
\mu_p\equiv \pm T_p\ ,~~~
\lambda\equiv 2\pi l_s^2\ ,
\label{T-mu-lambda}
\end{align}
where $l_s$ is the string length, $g_s$ is the string coupling,
and the upper (lower) sign appearing in $\mu_p$, which is proportional
to the R-R charge of the D$p$-brane, corresponds to the case of D$p$-branes
($\overline{\text{D}p}$-branes).
The quantities $M_{\mu\nu}$ and $Q^i_{~j}$ are given by
\begin{eqnarray}
M_{\mu\nu}
&\equiv&{\rm P}\left[
\wh E_{\mu\nu}+\wh E_{\mu i}(Q^{-1}-\delta)^{ij}\wh E_{j\nu}
\right]+
\lambda F_{\mu\nu}\ ,
\label{Malphabeta}
\\
Q^i_{~j}&\equiv&\delta^i_{~j}+i \lambda [\Phi^i,\Phi^k]\wh E_{kj}\ ,
 \label{Qij}
\end{eqnarray}
where
$F$ is the field strength of the gauge field $A$ living on the brane,
\begin{align}
F = dA+ iA\wedge A=\half F_{\mu\nu}dx^\mu\wedge dx^\nu
\ ,
\end{align}
and
\begin{eqnarray}
\wh E_{IJ}\equiv\wh g_{IJ}+\wh B_{IJ}\ .
\end{eqnarray}
$\phi$ is the dilaton field,
$g_{IJ}$ is the background metric, $B_2=\half B_{IJ}dx^I\wedge dx^J$
is the Kalb-Ramond 2-form field and $C_n$ ($n=0,2,4,6,8$)
are the Ramond-Ramond (R-R) $n$-form potential.
The hat ``\,$\wh{~}$\,'' on the background fields indicates that
they are evaluated at the position of the D$p$-branes
placed at $x^i=\lambda\Phi^i$, which is defined via a Taylor expansion
as, {\it e.g.},
\begin{align}
\wh\phi(x^\mu,\lambda\Phi^i)
\equiv
\sum_{n=0}^\infty\frac{\lambda^n}{n!} \Phi^{i_1} \dots \Phi^{i_n}
\left.\partial_{i_1}\dots \partial_{i_n}\phi(x^\mu,x^i)\right|_{x^i=0}
\ .
\label{expand}
\end{align}

The symbol $\rm P[\cdots]$ in (\ref{CS-initial}) and (\ref{Malphabeta})
denotes the pull-back of the bulk
fields over the D$p$-brane world-volume, in which
the ordinary  derivative $\del_\mu\Phi^i$ is replaced by
the covariant derivative  $D_\mu\Phi^i$:
\begin{align}
D_\mu\Phi^i \equiv \partial_\mu\Phi^i+i[A_\mu,\Phi^i]
\ .
\label{delPhi}
\end{align}
For example, the pull-back of $E_{\mu\nu}$ is given by
\begin{eqnarray}
 {\rm P}[E_{\mu\nu}]=
\wh E_{\mu\nu}+\lambda\wh E_{\mu i}D_\nu\Phi^i
+\lambda\wh E_{i\nu}D_\mu\Phi^i
+\lambda^2\wh E_{ij}D_\mu\Phi^iD_\nu\Phi^j\ .
\label{PE}
\end{eqnarray}
$\imath_\Phi$ in (\ref{CS-initial}) denotes the interior product
by a vector $(\Phi^i)$, {\it e.g.},
\begin{eqnarray}
 \imath_\Phi \imath_\Phi
\left(\half C_{ij}dx^i\wedge dx^j\right)
=-\half C_{ij}[\Phi^i,\Phi^j]\ .
\label{iPhi}
\end{eqnarray}

The symbol ${\rm Str\{\cdots\}}$
in (\ref{DBI-initial}) and  (\ref{CS-initial})
 denotes the symmetrized trace,
which means $\Phi^i$ in the expansion (\ref{expand}), $F_{\mu\nu}$,
$D_\mu\Phi^i$ and $[\Phi^i,\Phi^j]$ are symmetrized
before taking the trace.

\subsubsection{Fermionic part}

In this subsection, we write down the fermionic part (quadratic terms
with respect to the fermion fields) of the effective action
on a D$p$-brane embedded in any supergravity background
following \cite{Martucci:2005rb}. Here, we consider the cases
with a single D$p$-brane in type IIB string theory.

The action, after fixing the $\kappa$-symmetry, is given by
\begin{multline}
 S_{{\rm D}p}^{\rm fermi}
=
\frac{T_{p}}{2}
\int d^{p+1}x\, e^{-\phi}
\sqrt{-\det(M_{\mu\nu})}\Big\{
i\ol\psi
\left[(M^{-1})^{\mu\nu}\Gamma_\mu
\nabla_\nu^{(H)}-\Delta^{(1)}\right]\psi
\\
-i\ol\psi\,\breve\Gamma_{Dp}^{-1}
\left[(M^{-1})^{\mu\nu}
\Gamma_\nu W_\mu-\Delta^{(2)}\right]\psi
\Big\}\ ,
\label{Sfermi}
\end{multline}
where $\psi$ is the fermion field (dimensional reduction
of the 10-dimensional positive chirality Majorana-Weyl spinor field),
$\Gamma_\mu\equiv \Gamma_{\hat I}e_I^{\hat I}\del_\mu x^I$
 is the pull-back of the 10-dimensional gamma matrices,
 $M_{\mu\nu}$ is the Abelian version of (\ref{Malphabeta}):
\begin{eqnarray}
 M_{\mu\nu}={\rm P}[\wh g_{\mu\nu}+\wh B_{\mu\nu}]+\lambda F_{\mu\nu}\ , 
\end{eqnarray}
and other quantities are defined as follows.

The covariant derivative $\nabla_\nu^{(H)}$ is the pull-back of the
10-dimensional covariant derivative including the $H$-flux:
\begin{eqnarray}
\nabla_I^{(H)}\equiv
 \del_I+\frac{1}{4}\omega_I^{~\hat J\hat K}
\Gamma_{\hat J\hat K}
+\frac{1}{4\cdot 2!} H_{IJK}\Gamma^{JK}
\ ,
\label{nablaH}
\end{eqnarray}
where $\omega_{I}^{~\hat J\hat K}$ is the spin connection
and $H_{IJK}$ is the field strength of the Kalb-Ramond 2-form
field.

$W_\mu$, $\Delta^{(1)}$ and $\Delta^{(2)}$ are
defined as
\begin{align}
W_\mu
\equiv&\
\frac{1}{8}e^{\phi}\left(
-F_{J}\Gamma^{J}
+\frac{1}{3!}\wt F_{JKL}\Gamma^{JKL}
-\frac{1}{2\cdot 5!}\wt F_{JKLMN}\Gamma^{JKLMN}
\right)\Gamma_\mu
\ ,
\label{Wmu}
\\
\Delta^{(1)}
\equiv&\
\half\left(
\Gamma^I\del_I\phi+\frac{1}{2\cdot 3!}H_{IJK}\Gamma^{IJK}
\right)
\ ,
\label{Delta1}
\\
\Delta^{(2)}
\equiv&\
-\frac{1}{2}e^\phi\left(
-F_{I}\Gamma^{I}+\frac{1}{2\cdot 3!}\wt F_{IJK}\Gamma^{IJK}
\right)
\ ,
\label{Delta2}
\end{align}
where $F_I$, $\wt F_{IJK}$ and $\wt F_{IJKLM}$
are the field strength of the R-R fields
defined as
\begin{eqnarray}
&&F_1=F_Idx^I\equiv dC_0\ ,~~~
\wt F_3=\frac{1}{3!}\wt F_{IJK}dx^I\wedge dx^J\wedge dx^K
\equiv dC_2+C_0 H_3
\ ,\nn\\
&&
\wt F_5
=\frac{1}{5!}\wt F_{IJKLM}
dx^I\wedge dx^J\wedge dx^K\wedge dx^L\wedge dx^M
\equiv dC_4+H_3\wedge C_2\ .
\end{eqnarray}
(See Appendix \ref{app:SUGRA} for our conventions.)

Finally, $\breve\Gamma_{\text{D}p}^{-1}$ is defined by
\begin{eqnarray}
\breve\Gamma_{\text{D}p}^{-1}
=(-1)^{p-2}\Gamma_{\text{D}p}^{(0)}
\frac{\sqrt{-g}}{\sqrt{-\det(M_{\mu\nu})}}
\sum_{q\ge 0} \frac{(-1)^q}{q!2^q}\Gamma^{\mu_1\cdots\mu_{2q}}
\cF_{\mu_1\mu_2}\cdots \cF_{\mu_{2q-1}\mu_{2q}}\ ,
\end{eqnarray}
where $\sqrt{-g}\equiv\sqrt{-\det{\rm P}[\hat g_{\mu\nu}]}$\ ,
\begin{eqnarray}
 \cF_{\mu\nu}\equiv {\rm P}[\wh B_{\mu\nu}]+\lambda F_{\mu\nu}\ ,
\end{eqnarray}
and
\begin{eqnarray}
 \Gamma_{\text{D}p}^{(0)}\equiv
\frac{1}{(p+1)!}\epsilon^{\mu_1\cdots\mu_{p+1}}
\Gamma_{\mu_1\cdots\mu_{p+1}}
\end{eqnarray}
with the Levi-Civita symbol $\epsilon^{\mu_1\cdots\mu_{p+1}}$
with $\epsilon^{01\cdots p}=1/\sqrt{-g}$.

\subsection{D3-brane effective action}
\label{app:D3-ourcase}

In this appendix,
we consider the particular case of D$3$-branes
under some simple and relatively general assumptions (\ref{assumptions}).
We will study the
expansion of the full action to leading and sub-leading orders that
survive in the field theory limit
and establish a relation between the backgrounds fields and the couplings
in the action  (\ref{action1}) of the deformed $\cN=4$ SYM.
The first two subsections  \ref{app:DBI} and \ref{app:CS} correspond to
the analyses for the DBI action and the CS term, respectively, and
Appendix \ref{app:bosonic} is the summary of the total bosonic sector.
In Appendix \ref{app:fermionic}, we carry out the calculations
for the fermionic sector.

\subsubsection{DBI action}
\label{app:DBI}
In this section, we present the extended calculations of the expansion
of the DBI term \eqref{DBI-initial} with respect to $\lambda$.
Let us first consider the quantity $M_{\mu\nu}$ defined in
\eqref{Malphabeta}.
The pull-back of the first term of (\ref{Malphabeta}) is given in (\ref{PE})
and it is expanded as
\begin{align}	
{\rm P}[\wh E_{\mu\nu}]
=&\
\wh E_{\mu\nu}
+\lambda \wh E_{\mu i}D_\nu\Phi^i
+\lambda \wh E_{i\nu}D_\mu\Phi^i
+\lambda^2\wh E_{ij}D_\mu\Phi^i D_\nu\Phi^j
\nn\\
=&\
\wh E_{\mu\nu}+\lambda^2 (\del_jB_{\mu i}\Phi^jD_\nu\Phi^i
+\del_jB_{i\nu}\Phi^jD_\mu\Phi^i
+g_{ij}D_\mu\Phi^iD_\nu\Phi^j)+\cO(\lambda^3)\ ,
\end{align}
where we have used the assumptions
in (\ref{assumptions}).
The expansion of $Q^i{}_j$ in (\ref{Qij}) is
\begin{align}
Q^i_{~j}
=&\ \delta^i_{~j}+i\lambda[\Phi^i,\Phi^k]g_{kj}
+i\lambda^2[\Phi^i,\Phi^k]\Phi^l\del_l E_{kj}+\cO(\lambda^3)
\ .
\end{align}
Because $(Q^{-1}-\delta)^i_j=\cO(\lambda)$,
it is easy to see that
\begin{eqnarray}
{\rm P}\left[E_{\mu i}(Q^{-1}-\delta)^{ij}E_{j\nu}\right]
 =\cO(\lambda^3)\ ,
\end{eqnarray}
under the assumptions (\ref{assumptions})
and we can discard such higher-order contributions.

On the other hand, using the formula
\begin{multline}
\sqrt{\det(X+\delta X)}
=
\sqrt{\det X}\left(1+\half\tr(X^{-1}\delta X)
+\frac{1}{8}(\tr(X^{-1}\delta X))^2
\right.
\\
\left.
-\frac{1}{4}\tr(X^{-1}\delta X X^{-1}\delta X)+\cO(\delta X^3)
\right)
\end{multline}
for general matrices $X$ and $\delta X$, we get
\begin{align}
\sqrt{\det(Q^i{}_j)}
=1+\frac{i\lambda^2}{2}[\Phi^i,\Phi^k]\Phi^l\del_lB_{ki}
-\frac{\lambda^2}{4}g_{ii'}g_{jj'}[\Phi^i,\Phi^j][\Phi^{i'},\Phi^{j'}]
+\cO(\lambda^3)\ .
\label{detQ}
\end{align}

Similarly, $\sqrt{-\det(M_{\mu\nu})}$ is expanded as
\begin{align}
\sqrt{-\det(M_{\mu\nu})}
=&\
\sqrt{-\det (\wh E_{\mu\nu})}
\nn\\
&\
+\sqrt{-g}\
\frac{\lambda^2}{2}
\Big(
g^{\mu\nu}g_{ij}D_\mu\Phi^iD_\nu\Phi^j
+\half g^{\mu\nu}g^{\rho\sigma}F_{\mu\rho}F_{\nu\sigma}
+(\del_i B^{\mu\nu})\Phi^iF_{\mu\nu}
\Big)
+\cO(\lambda^3)
\ .
\label{detM}
\end{align}

Now, making use of the partial results of the expansions \eqref{detQ} and
\eqref{detM}, we calculate the full integrand of the DBI term (\ref{DBI-initial}):
\begin{align}
{\rm Str}
\left\{e^{-\wh\phi}
\sqrt{-\det(M_{\mu\nu})\det(Q^i_{~j})}
\right\}
\hspace{-5cm}&
\nn\\
=&\
 {\rm Str}\left\{e^{-\wh\phi}
\sqrt{-\det(\wh E_{\mu\nu})}\right\}
\nn\\
&\
+\frac{\lambda^2}{2}
e^{-\phi}
\sqrt{-g}
\tr\bigg(
\frac{1}{2}g^{\mu\nu}g^{\rho\sigma}F_{\mu\rho}F_{\nu\sigma}
+g^{\mu\nu}g_{ij}D_{\mu}\Phi^iD_{\nu}\Phi^j
-\frac{1}{2}g_{ii'}g_{jj'}[\Phi^i,\Phi^j][\Phi^{i'},\Phi^{j'}]
\nn\\
&\
+(\del_iB^{\mu\nu})\Phi^iF_{\mu\nu}
-i(\del_iB_{jk})\Phi^i[\Phi^j,\Phi^k]
\bigg)
+\cO(\lambda^3)
\nn\\
=&\ {\rm Str}\left\{e^{-\wh\phi}
\sqrt{-\det(\wh E_{\mu\nu})}\right\}
\nn\\
&\
+\frac{\lambda^2}{2}
e^{-\phi}
\sqrt{-g}
\tr\bigg(
\half
 g^{\mu\nu}g^{\rho\sigma}F_{\mu\rho}F_{\nu\sigma}
+g^{\mu\nu}g_{ij}D_\mu\Phi^i D_\nu\Phi^j
-\half
g_{ii'}g_{jj'}[\Phi^i,\Phi^j][\Phi^{i'},\Phi^{j'}]
\bigg)
\nn\\
&\
+H_{i}^{~\mu\nu}
\Phi^iF_{\mu\nu}
-\frac{i}{3}H_{ijk}\Phi^i[\Phi^j,\Phi^k]
\bigg)+\cO(\lambda^3)\ .
\label{DBIexp}
\end{align}
The first term in (\ref{DBIexp}) gives the DBI part of the scalar
potential. Let us define
\begin{eqnarray}
 V_{\rm DBI}(\Phi)\equiv\frac{1}{\sqrt{-g}}
e^{-\wh\phi}\sqrt{-\det(\wh E_{\mu\nu})}
=\frac{1}{\sqrt{-g}}\sum_{n=0}^\infty
\frac{\lambda^n}{n!}\Phi^{i_1}\cdots\Phi^{i_n}
\del_{i_1}\cdots\del_{i_n}L_{\rm DBI}|_{x^i=0}
\end{eqnarray}
with
\begin{align}
L_{\rm DBI}\equiv e^{-\phi}\sqrt{-\det(E_{\mu\nu})}\ .
\label{LDBI}
\end{align}
The derivatives are
\begin{align}
\del_i L_{\rm DBI}
=&\ 
e^{-\phi}\sqrt{-\det(E_{\mu\nu})}
\left(
-\del_i\phi+\half (E^{-1})^{\mu\nu}\del_iE_{\nu\mu}
\right)
\ ,
\\
\del_i\del_j L_{\rm DBI}
=&\ 
e^{-\phi}\sqrt{-\det(E_{\mu\nu})}
\left[
\left(-\del_i\phi+\half (E^{-1})^{\mu\nu}\del_iE_{\nu\mu}\right)
\left(-\del_j\phi+\half (E^{-1})^{\mu'\nu'}\del_jE_{\nu'\mu'}\right)
\right.
\nn\\
&\
\left.
\quad\qquad
-\del_i\del_j\phi+\half\left(
-(E^{-1})^{\mu\mu'}\del_i E_{\mu'\nu'}(E^{-1})^{\nu'\nu}\del_j E_{\nu\mu}
+(E^{-1})^{\mu\nu}\del_i\del_j E_{\nu\mu}
\right)
\right]
\ .
\end{align}
Evaluating these quantities at $x^i=0$, we obtain
\begin{eqnarray}
V_{\rm DBI}(\Phi)
=
e^{-\phi}\left(
1+\lambda
v^{\rm DBI}_i\Phi^i
+\frac{\lambda^2}{2}m^{\rm DBI}_{ij}
\Phi^i\Phi^j
+\cO(\lambda^3)
\right)
\ ,
\end{eqnarray}
where the coefficients are
\begin{align}
v^{\rm DBI}_i
\equiv&\
 -\del_i\phi+\frac{1}{2}g^{\mu\nu}\del_ig_{\nu\mu}
=\del_i\log(\sqrt{-g}\,e^{-\phi})
\ ,
\\
m^{\rm DBI}_{ij}
\equiv&\
v_i^{\rm DBI}v_j^{\rm DBI}
-\del_i\del_j\phi+\half\left(
g^{\mu\nu}\del_i\del_j g_{\nu\mu}
-g^{\mu\mu'}\del_i g_{\mu'\nu'} g^{\nu'\nu}\del_j g_{\nu\mu}
-g^{\mu\mu'}H_{i\mu'\nu'} g^{\nu'\nu}H_{j\nu\mu}
\right)
\nn\\
=&\ \frac{1}{\sqrt{-g}\,e^{-\phi}}\del_i\del_j
\left(\sqrt{-g}\,e^{-\phi}\right)+\half H_{i}^{~\mu\nu} H_{j\mu\nu}
\ .
\end{align}

Then, the final expression for the DBI action is
\begin{align}
S_{\rm D3}^{\rm DBI}
=&\
T_3\lambda^2\int d^4 x
\sqrt{-g}\,e^{-\phi}
\tr\bigg(
-\frac{1}{4}g^{\mu\nu}g^{\rho\sigma}F_{\mu\rho}F_{\nu\sigma}
-\frac{1}{2}g^{\mu\nu}g_{ij}D_{\mu}\Phi^iD_{\nu}\Phi^j
\nn\\
&\
+\frac{1}{4}g_{ii'}g_{jj'}[\Phi^i,\Phi^j][\Phi^{i'},\Phi^{j'}]
-\frac{1}{2}H_{i}^{~\mu\nu}\Phi^iF_{\mu\nu}
+\frac{i}{3!}H_{ijk}\Phi^i[\Phi^j,\Phi^k]
-e^\phi\lambda^{-2}V_{\rm DBI}(\Phi)
\bigg)\ .
\label{DBID3}
\end{align}

\subsubsection{CS term}
\label{app:CS}

Let us study now the expansion of the CS-term of the D3-brane action,
\eqref{CS-initial}.

First, we define
\begin{eqnarray}
K\equiv \sum_{n:{\rm even}}C_n\wedge e^{B_2}
\equiv K_0+K_2+K_4+K_6+\cdots\ ,
\end{eqnarray}
where\footnote{In this section, we often omit the symbol ``$\wedge$''
in the products of differential forms.}
\begin{eqnarray}
&&
K_0\equiv C_0\ ,~~~
K_2\equiv C_2+C_0B_2\ ,~~~
K_4\equiv C_4+C_2B_2+\half C_0B_2^2\ ,
\nn\\
&&K_6\equiv C_6+C_4B_2+\half C_2B_2^2+\frac{1}{3!}C_0B_2^3\ .
\end{eqnarray}
Note that it satisfies
\begin{eqnarray}
dK=\sum_{n:{\rm odd}} \wt F_n\wedge e^{B_2}
\ .
\label{dKF}
\end{eqnarray}

For an $n$-form $\omega_n$, we define an $m$-form with $(n-m)$
indices in the transverse directions $(\omega_n)_{m,i_1\cdots i_{n-m}}$ as
\begin{eqnarray}
(\omega_n)_{m,i_1\cdots i_{n-m}}
\equiv\frac{1}{m!} (\omega_n)_{\mu_1\cdots\mu_m i_1\cdots i_{n-m}}
dx^{\mu_1}\cdots dx^{\mu_m}\ .
\label{omega-nm}
\end{eqnarray}
For example,
\begin{eqnarray}
(\wt F_5)_{3,ij}\equiv \frac{1}{3!}(\wt F_5)_{\mu\nu\rho ij}
dx^\mu dx^\nu dx^\rho\ ,~~~
 (\wt F_5)_{4,j}\equiv \frac{1}{4!}\wt F_{\mu\nu\rho\sigma j}
dx^\mu dx^\nu dx^\rho dx^\sigma\ ,~~~\mbox{etc.}
\end{eqnarray}

Under the assumptions \eqref{assumptions}, the CS-term \eqref{CS-initial}
is expanded as
\begin{align}
 S_{{\rm D}3}^{\rm CS}
=&\
\mu_{3}\int
{\rm Str}\left\{{\rm P}
\left[e^{i\lambda \imath_\Phi\imath_\Phi}
\wh K
\right]
\wedge e^{\lambda F}
\right\}
\nn\\
=&\
\mu_{3}\int
{\rm Str}\bigg\{
{\rm P}[\wh K_4]
+\lambda^2\left(
i{\rm P}[\imath_\Phi\imath_\Phi
\del_i K_6]\Phi^i
+{\rm P}[\del_i K_2]\Phi^i F
+\frac{1}{2}C_0F^2
\right)
+\cO(\lambda^3)\bigg\}\ .
\label{CS-1}
\end{align}
Expanding the first term, we get
\begin{align}
{\rm P}[\wh K_4]
=&\
\left[\lambda \del_i K_4\Phi^i+\lambda^2 \del_i(K_4)_{3,j}\Phi^i D\Phi^j
+\frac{\lambda^2}{2}\del_i\del_j K_4\Phi^i\Phi^j\right]_0
+\cO(\lambda^3)\ ,
\label{PK4}
\end{align}
where $[\cdots]_0$ denotes the pull-back on the world-volume
at $x^i=0$ (obtained by setting $x^i=0$ and $dx^i=0$),
 $D\Phi^j$ is a 1-form defined as
\begin{align}
 D\Phi^j\equiv&\
D_\mu \Phi^j dx^\mu=d\Phi^j+i[A,\Phi^j]\ ,
\end{align}
and we have used the notation (\ref{omega-nm}).

The trace of the second term in (\ref{PK4}) can be rewritten as
\begin{align}
&\
\left[ \lambda^2\del_i(K_4)_{3,j}\tr(\Phi^iD\Phi^j)\right]_0
\nn\\
~~~
=&\
\frac{\lambda^2}{2}
\left[\half(\del_i(K_4)_{3,j}+\del_j(K_4)_{3,i})d\tr(\Phi^i\Phi^j)
+(\del_i(K_4)_{3,j}-\del_j(K_4)_{3,i})\tr(\Phi^iD\Phi^j)\right]_0
\nn\\
=&\
\frac{\lambda^2}{2}
\left[(\del_id(K_4)_{3,j}\tr(\Phi^i\Phi^j)
-(\wt F_5)_{3,ij}\tr(\Phi^iD\Phi^j)+(\mbox{total derivative})
\right]_0\ .
\end{align}
Using this equation and the identities
\begin{eqnarray}
&&[\del_i K_4]_0=[(\wt F_5)_{4,i}]_0\ ,
\nn\\
&&
[\del_id(K_4)_{3,j}+\del_i\del_j K_4]_0
=[\del_i(dK_4)_{4,j}]_0
=[\del_i(\wt F_5)_{4,j}+(\wt F_3)_{2,j}(H_3)_{2,i}]_0\ ,
\end{eqnarray}
which are valid under the assumptions (\ref{assumptions}), we obtain
\begin{align}
\int {\rm Str}\,{\rm P}[\wh K_4]
=
\int\tr\left\{
\lambda(\wt F_5)_{4,i}\Phi^i
+\frac{\lambda^2}{2}
\left(\del_i(\wt F_5)_{4,j}+(\wt F_3)_{2,j}(H_3)_{2,i}\right)
\Phi^i\Phi^j
-\frac{\lambda^2}{2}
(\wt F_5)_{3,ij}\Phi^iD\Phi^j
\right\}\ .
\end{align}

The second and third terms in (\ref{CS-1}) are
rewritten by using
\begin{align}
\tr\{i{\rm P}[\imath_\Phi\imath_\Phi \del_iK_6]\Phi^i\}
=&\ -\frac{i}{2}[\del_i(K_6)_{4,jk}]_0
\tr\{[\Phi^j,\Phi^k]\Phi^i\}
=-\frac{i}{3!}[(\wt F_7)_{4,ijk}]_0
\tr\{[\Phi^j,\Phi^k]\Phi^i\}
\ ,
\\
{\rm P}[\del_iK_2]\Phi^i F_2
=&\ [(\wt F_3)_{2,i}]_0\, \Phi^iF
\ ,
\end{align}
respectively.

Plugging these results in (\ref{CS-1}), the CS-term becomes
\begin{align}
 S_{\rm D3}^{\rm CS}
=&\
\mu_3\lambda^2
\int\tr\bigg\{
\lambda^{-1}(\wt F_5)_{4,i}\Phi^i
+\frac{1}{2}
\left(\del_i(\wt F_5)_{4,j}+(\wt F_3)_{2,j}(H_3)_{2,i}\right)
\Phi^i\Phi^j
-\frac{1}{2}
(\wt F_5)_{3,ij}\Phi^iD\Phi^j
\nn\\
&~~~~~~~~
-\frac{i}{3!}(\wt F_7)_{4,ijk}\Phi^i[\Phi^j,\Phi^k]
+(\wt F_3)_{2,i}\Phi^iF+\half C_0F^2
\bigg\}\ .
\label{CS-tmp}
\end{align}
It can also be written as
\begin{align}
 S_{\rm D3}^{\rm CS}
=&\
\mu_3\lambda^2
\int d^4x\sqrt{-g}
\tr\bigg\{
\half(*_4\wt F_3)^{~\mu\nu}_{i}\Phi^iF_{\mu\nu}
-\frac{i}{3!}(*_6\wt F_3)_{ijk}\Phi^i[\Phi^j,\Phi^k]
\nn\\
&~~~~~~~~
-\frac{1}{2}
(*_4\wt F_5)^\mu_{~ij}\Phi^iD_\mu\Phi^j
+\frac{1}{8} C_0F_{\mu\nu} F_{\rho\sigma}\epsilon^{\mu\nu\rho\sigma}
-\lambda^{-2}V_{\rm CS}(\Phi)
\bigg\}\ ,
\label{CSD3}
\end{align}
where the potential $V_{\rm CS}(\Phi)$ is
\begin{eqnarray}
 V_{\rm CS}(\Phi)\equiv
-\lambda(*_4\wt F_5)_{i}\Phi^i
-\frac{\lambda^2}{2}
\left(\del_i(*_4\wt F_5)_{j}+\half(*_4\wt F_3)^{~\mu\nu}_{j}(H_3)_{\mu\nu i}\right)
\Phi^i\Phi^j
\end{eqnarray}
and we have defined
\begin{eqnarray}
&& (*_4\wt F_3)_i^{~\mu\nu}
\equiv \frac{1}{2}\epsilon^{\rho\sigma\mu\nu}\wt F_{i\rho\sigma}\ ,
~~~
 (*_6\wt F_3)_{ijk}
\equiv \frac{1}{3!}\epsilon_{lmn ijk}\wt F^{lmn}\ ,
\nn\\
&&
(*_4\wt F_5)^\mu_{~ij}
\equiv \frac{1}{3!}\epsilon^{\nu\rho\sigma\mu}
\wt F_{\nu\rho\sigma ij}\ ,
~~~
 (*_4\wt F_5)_{i}
\equiv \frac{1}{4!}\epsilon^{\mu\nu\rho\sigma}
\wt F_{\mu\nu\rho\sigma i}\ ,
\label{wtF}
\end{eqnarray}
and used the relation
\begin{eqnarray}
 \wt F_7=-*\wt F_3\ .
\end{eqnarray}

\subsubsection{Bosonic part}
\label{app:bosonic}

Summing (\ref{DBID3}) and (\ref{CSD3}), we obtain
\begin{multline}
S_{\rm D3}^{\rm boson}
=
\frac{T_3\lambda^2}{2}\int d^4 x
\sqrt{-g}
\tr\bigg(
-\frac{e^{-\phi}}{2}g^{\mu\nu}g^{\rho\sigma}F_{\mu\rho}F_{\nu\sigma}
\pm\frac{C_0}{4}
\epsilon^{\mu\nu\rho\sigma}F_{\mu\nu}F_{\rho\sigma}
\\
-e^{-\phi}g^{\mu\nu}g_{ij}D_{\mu}\Phi^iD_{\nu}\Phi^j
+\frac{e^{-\phi}}{2}g_{ii'}g_{jj'}[\Phi^i,\Phi^j][\Phi^{i'},\Phi^{j'}]
-2V(\Phi)
\\
\pm (G^R_\pm)_i^{~\mu\nu}
\Phi^iF_{\mu\nu}
\mp\frac{i}{3}(G^R_\pm)_{ijk}\Phi^i[\Phi^j,\Phi^k]
\mp (*_4\wt F_5)^\mu_{~ij}\Phi^iD_\mu\Phi^j
\bigg)\ ,
\end{multline}
where we have set
\begin{eqnarray}
 \mu_3=\pm T_3
\end{eqnarray}
for D3-branes and $\overline{\text{D}3}$-branes, respectively, and defined
\begin{eqnarray}
G_3&\equiv& F_3+(C_0+i e^{-\phi}) H_3
 =\wt F_3+i e^{-\phi} H_3\ ,
\\
 (G^R_\pm)_i^{~\mu\nu}
&\equiv&{\rm Re}( (*_4 G_3)_i^{~\mu\nu}\pm i G_i^{~\mu\nu})
= (*_4 \wt F_3)_i^{~\mu\nu}\mp e^{-\phi} H_i^{~\mu\nu}
\ ,
\label{GR1}
\\
 (G^R_\pm)_{ijk}&\equiv&{\rm Re}((*_6 G_3)_{ijk}\pm i G_{ijk})
= (*_6 \wt F_3)_{ijk}\mp e^{-\phi} H_{ijk}
\ ,
\label{GR2}
\end{eqnarray}
and
\begin{eqnarray}
 V(\Phi)\equiv\lambda^{-2}(V_{\rm DBI}(\Phi)\pm V_{\rm CS}(\Phi))\ .
\end{eqnarray}

\subsubsection{Fermionic part}
\label{app:fermionic}

Let us consider the fermionic action (\ref{Sfermi}) for a D3-brane.
To make the kinetic term of the fermions $\cO(\lambda^0)$,
we rescale the fermion as
\begin{align}
\psi=\lambda\Psi\ .
\end{align}
Since we are interested in the terms that survive in the $l_s\ra 0$ limit,
we can set  $M_{\mu\nu}=g_{\mu\nu}$ and
$\breve\Gamma_{\text{D}3}^{-1}=-\Gamma_{{\rm D}3}^{(0)}
=\Gamma^{(4)}$ where
\begin{eqnarray}
 \Gamma^{(4)}\equiv\Gamma^{\hat 0}\Gamma^{\hat 1}\Gamma^{\hat 2}\Gamma^{\hat 3}\ .
\end{eqnarray}
Inserting (\ref{nablaH})-(\ref{Delta2}) into the action (\ref{Sfermi})
we obtain
\begin{align}
S_{\rm D3}^{\rm fermi}
=&\
\frac{T_3\lambda^2}{2}\int d^4 x\,\sqrt{-g}\, e^{-\phi}
\,
i\ol\Psi\bigg[
\Gamma^\mu\nabla_\mu
+\frac{1}{4\cdot 2!}H_{\mu IJ}\Gamma^{\mu IJ}
-\frac{1}{4\cdot 3!}H_{IJK}\Gamma^{IJK}
\nn\\
&\
\qquad\qquad\qquad
\mp e^\phi\Gamma^{(4)}
\bigg(
\frac{1}{8}\Gamma^\mu
\left(
-F_I\Gamma^I+\frac{1}{3!}\wt F_{IJK}\Gamma^{IJK}
-\frac{1}{2\cdot 5!}\wt F_{IJKL}\Gamma^{IJKL}
\right)\Gamma_\mu
\nn\\
&\
\qquad\qquad\qquad\qquad\qquad\qquad
-\frac12 F_I\Gamma^I+\frac{1}{4\cdot 3!}\wt F_{IJK}\Gamma^{IJK}
\bigg)
\bigg]\Psi
\ ,
\end{align}
where
 $\nabla_\mu$ is
\begin{eqnarray}
\nabla_\mu\Psi\equiv \del_\mu\Psi+
\frac{1}{4}\omega_\mu^{~\hat I\hat J}\Gamma_{\hat I\hat J}\Psi\ .
\end{eqnarray}
Again, the upper (lower) signs correspond to the case of 
D3- ($\overline{\text{D}3}$-) branes.
Note that the first term in (\ref{Delta1}) does not contribute, because
$\Gamma^0\Gamma^I$ is a symmetric matrix. In general, one can show
\begin{eqnarray}
\ol\Psi\,\Gamma^{I_1\dots I_n}\Psi=0 ~~~ \text{for }~~ n\ne 3
~ (\mod 4)\ .
\label{PsiGamPsi}
\end{eqnarray}
Using this fact and the identities:
\begin{align}
\Gamma^\mu(F_I\Gamma^I)\Gamma_\mu
=&\
-2 F_\mu\Gamma^\mu -4 F_i\Gamma^i
\ ,
\\
\Gamma^\mu(\wt F_{IJK}\Gamma^{IJK})\Gamma_\mu
=&\
2\wt F_{\mu\nu\rho}\Gamma^{\mu\nu\rho}
-6\wt F_{\mu jk}\Gamma^{\mu jk}
-4\wt F_{ijk}\Gamma^{ijk}
\ ,
\\
\Gamma^\mu(\wt F_{IJKLM}\Gamma^{IJKLM})\Gamma_\mu
=&\
20\wt F_{\mu\nu\rho\sigma m}\Gamma^{\mu\nu\rho\sigma m}
+20\wt F_{\mu\nu\rho lm}\Gamma^{\mu\nu\rho lm}
-10\wt F_{\mu jklm}\Gamma^{\mu jklm}
-4\wt F_{ijklm}\Gamma^{ijklm}\ ,
\end{align}
we obtain
\begin{align}
S_{\rm D3}^{\rm fermi}
=&\
\frac{T_3\lambda^2}{2}\int d^4 x\, \sqrt{-g}\,e^{-\phi}
\,
i\ol\Psi\bigg[
\Gamma^\mu\nabla_\mu
+\frac{1}{4\cdot 3!}(
3H_{i\mu\nu}\Gamma^{i\mu\nu}-H_{ijk}\Gamma^{ijk})
\nn\\
&\
\qquad
\mp e^\phi\Gamma^{(4)}
\bigg(-\frac{1}{4}F_\mu\Gamma^\mu
+\frac{1}{4\cdot 3!}(3\wt F_{i\mu\nu}\Gamma^{i\mu\nu}
-\wt F_{ijk}\Gamma^{ijk})
\nn\\
&\
\qquad\qquad\qquad\qquad
-\frac{1}{3!\cdot 2^5}(
2\wt F_{\mu\nu\rho ij}\Gamma^{\mu\nu\rho ij}
-\wt F_{\mu ijkl}\Gamma^{\mu ijkl}
)\bigg)\bigg]\Psi
\ .
\end{align}

Furthermore, using the following identities
\begin{align}
&\Gamma^{(4)}\Gamma^{\mu}=
\frac{1}{3!}\epsilon^{\mu\nu\rho\sigma}\Gamma_{\nu\rho\sigma}
\ ,~~~
\Gamma^{(4)}\Gamma^{\mu\nu}=
\frac{1}{2}\epsilon^{\mu\nu\rho\sigma}\Gamma_{\rho\sigma}
\ ,~~~
\Gamma^{(4)}\Gamma^{\mu\nu\rho}=
-\epsilon^{\mu\nu\rho\sigma}\Gamma_{\sigma}
\ ,~~~
\nn\\
&\Gamma^{(4)}\Gamma^{ijk}=
\frac{1}{3!} \epsilon^{ijklmn}\Gamma_{lmn}\Gamma^{(10)}
\ ,~~~
\Gamma^{(4)}\Gamma^{ijkl}=
-\frac{1}{2} \epsilon^{ijklmn}\Gamma_{mn}\Gamma^{(10)}
\ ,
\end{align}
together with the chirality condition (\ref{chiralcond})
and the relation
\begin{eqnarray}
(*_4\wt F_5)_{\mu ij}=(*_6\wt F_5)_{\mu ij}\ ,
\end{eqnarray}
that follows from the self-duality condition (\ref{F5selfdual}),
the action can be rewritten as
\begin{align}
S_{\rm D3}^{\rm fermi}
=&\
\frac{T_3\lambda^2}{2}\int d^4 x\, \sqrt{-g}\,e^{-\phi}
\,i\ol\Psi\bigg[
\Gamma^\mu\nabla_\mu
+\frac{1}{4\cdot 3!}(
3H_{i\mu\nu}\Gamma^{i\mu\nu}-H_{ijk}\Gamma^{ijk})
\nn\\
&\
\mp e^\phi
\bigg(
-\frac{1}{4\cdot 3!}(*_4F_1)_{\mu\nu\rho}\Gamma^{\mu\nu\rho}
+\frac{1}{4\cdot 3!}\left(3(*_4\wt F_3)_{i\mu\nu}\Gamma^{i\mu\nu}
-(*_6\wt F_3)_{ijk}\Gamma^{ijk}\right)
+\frac{1}{8}(*_4\wt F_5)_{\mu ij}\Gamma^{\mu ij}
\bigg)\bigg]\Psi
\ ,
\end{align}
where we have used the notation (\ref{wtF}) and defined
\begin{eqnarray}
  (*_4F_1)_{\nu\rho\sigma}
\equiv
 \epsilon_{\mu\nu\rho\sigma}F^\mu
= \epsilon^{\mu}_{~\nu\rho\sigma}\del_\mu C_0\ .
\label{F1star}
\end{eqnarray}

For the non-Abelian case, the covariant derivative $\nabla_\mu$
should be replaced with
\begin{eqnarray}
\Gamma^\mu\nabla_\mu \Psi
\quad\ra \quad
\Gamma^\mu D_\mu\Psi+i\Gamma_k[\Phi^k,\Psi]
+\frac{1}{4}\omega_{\mu\hat i\hat j}\Gamma^{\mu\hat i\hat j}
\ ,
\end{eqnarray}
where $D_\mu\Psi$ is defined in (\ref{delPsi}) and
we have assumed $g_{\mu i}=0$ and $\omega_{\mu\hat \nu\hat i}=
-\omega_{\mu\hat i\hat \nu}=0$.

Then, our final expression for the fermionic part of the action is
\begin{align}
S_{\rm D3}^{\rm fermi}
=
\frac{T_3\lambda^2}{2}\int d^4 x\, \sqrt{-g}\,e^{-\phi}
\tr\left\{
i(\ol\Psi\Gamma^\mu D_\mu\Psi
+\ol\Psi\Gamma_k i[\Phi^k,\Psi])
-i\ol\Psi\left( M_\pm
-\frac{1}{4}\omega_{\mu\hat i\hat j}\Gamma^{\mu\hat i\hat j}
\right) \Psi \right\}
\end{align}
with
\begin{align}
M_\pm
\equiv&\
 \mp\frac{e^{\phi}}{8}\left(
\frac{1}{3}(*_4 F_1)_{\mu\nu\rho}\Gamma^{\mu\nu\rho}
- (G^R_\pm)_{i\mu\nu}\Gamma^{i\mu\nu}
+\frac{1}{3}(G_\pm^R)_{ijk}\Gamma^{ijk}
-(*_4\wt F_5)_{\mu ij}\Gamma^{\mu ij}
\right)
\ ,
\end{align}
where $(G_\pm^R)_{i\mu\nu}$ and $(G_\pm^R)_{ijk}$ are defined in (\ref{GR1})
and (\ref{GR2}), respectively.

\providecommand{\href}[2]{#2}\begingroup\raggedright\endgroup

\end{document}